\documentclass[a4paper,11pt]{article}
\pdfoutput=1 

\usepackage{jheppub} 

\usepackage[T1]{fontenc} 

\usepackage{subfigure}
\usepackage{braket}
\usepackage[compat=1.1.0]{tikz-feynman}
\usepackage[bottom]{footmisc}
\usepackage{cleveref}
\newcommand{\AH}[1]{{\leavevmode\color{blue}{#1}}}

\title{\boldmath Freezing-in Cannibal Dark Sectors}

\author{Esau Cervantes}
\author{{\rm and} Andrzej Hryczuk}

\affiliation{National Centre for Nuclear Research,\\ Pasteura 7, 02-093 Warsaw, Poland}

\emailAdd{esau.cervantes@ncbj.gov.pl}
\emailAdd{andrzej.hryczuk@ncbj.gov.pl}

\abstract{Self-Interacting Dark Matter models can successfully explain dark matter (DM) production through interactions confined within the dark sector. However, they often lack measurable experimental signals due to their secluded nature. Including a feeble interaction with the visible sector through a Higgs portal leads not only to potential detection avenues and richer thermal production dynamics, but also to a possible explanation of the initial dark sector population through the freeze-in mechanism. In this work we study, by solving the full system of coupled Boltzmann equations for the number densities and temperatures of all the involved states, three scenarios of this type where the DM is: a real scalar with broken $\mathbb{Z}_2$, a complex scalar with unbroken $\mathbb{Z}_3$, and a $\mathbb{Z}_3$ scalar with an additional scalar mediator. All of these models have viable dark matter candidates in a cannibal phase while having different detection profiles. We show that cosmological bounds can be either exacerbated or evaded by changing the dark sector interactions, leading to potential signatures in long-lived particle and indirect detection experiments.}

\begin{document} 
\maketitle
\flushbottom

\section{Introduction}
\label{sec:intro}

The astrophysical and cosmological evidence supporting the existence of dark matter is well established~\cite{Bertone:2010zza}.  
However, despite a large number of theoretical ideas on how to explain the presence of DM and decades of efforts put into experimental searches its nature remains unknown. One of the leading candidates for the DM particle is the so-called WIMP (\textit{Weakly Interacting Massive Particle}) which elegantly accounts for the observed abundance ($\Omega_c h^2 = 0.120\pm 0.001$~\cite{Planck:2018vyg}) by postulating weak-scale interactions with visible matter, while at the same time preserving all the other essential properties requisite for a viable candidate (see e.g.~\cite{Feng:2022rxt,Arcadi:2024ukq}). Nevertheless, lack of an experimental identification of any definitive WIMP signature motivated diverse models to elucidate the nature of DM without reliance on sizable interactions with visible matter. For instance sterile neutrinos~\cite{Boyarsky:2018tvu}, axions~\cite{OHare:2024nmr} or fuzzy dark matter~\cite{Anchordoqui:2023tln}. Another such alternative is the \textit{self-interacting DM} (SIDM) paradigm, where the DM abundance can be set through a freeze-out occurring within the dark sector due to DM self-number changing reactions, mechanism originally proposed in~\cite{1992ApJ...398...43C}. Interestingly, the freeze-out of self-number changing reactions (which we will refer to as ``dark freeze-out'') is characterized by a \textit{cannibalization} phase in which the dark sector converts its rest mass into kinetic energy. During this period DM is hotter than in the standard WIMP freeze-out scenario, which may potentially erase small-scale structures in the Universe leading to disagreement with observations. This phenomenon was first acknowledged in~\cite{1992ApJ...398...43C} and has been further discussed in the subsequent literature~\cite{deLaix:1995vi,Irsic:2017ixq,Chatterjee:2019olf,Buen-Abad:2018mas,Erickcek:2020wzd,Heimersheim:2020aoc}. To overcome these constraints, one strategy involves considering a dark sector which is initially colder than the visible one, so that despite warming during dark freeze-out, the dark sector remains sufficiently cold for successful structure formation~\cite{Hufnagel:2022aiz,Farina:2016llk,Ghosh:2022asg,Arcadi:2019oxh}. Another possibility relies on postulating a weak portal to visible matter, allowing equilibriation and transfer heat with the SM plasma during dark freeze-out, a scenario referred to as the ``SIMP miracle'' introduced in ~\cite{Hochberg:2014dra,Hochberg:2014kqa} and subsequently discussed in~\cite{Bernal:2015bla,Kuflik:2015isi,Pappadopulo:2016pkp,Kuflik:2017iqs}. Variations of this idea involve a DM candidate with an unstable Higgs-like mediator, leading to the depletion of dark sector's energy density during cannibalization due to the mediator's decay~\cite{Yang:2023xgk}. 

In both cases the initial production leading to a thermal population of particles of the dark sector in the first place is typically taken for granted. Among the mechanisms proposed to explain the origin of DM are gravitational production~\cite{Chung:2001cb}, inflaton decay~\cite{Takahashi:2007tz,Moroi:2020has}, asymmetric reheating~\cite{Feng:2008mu}, decay of false vacua~\cite{Asadi:2021pwo} and freeze-in (FI). The FI mechanism, in particular, relies on feeble interactions with visible matter~\cite{Lebedev:2019ton}, which are in fact typically present if the dark sector contains a scalar field that would naturally couple to the Higgs. This mechanism relies on three basic assumptions: 1) the initial abundance of the dark sector after reheating ends is zero or negligible;  2) any portal to matter is sufficiently weak, ensuring that DM is produced from the SM plasma but never thermalizes with it; and 3) the interaction populating the dark sector is renormalizable making the mechanism independent of the reheating temperature~\cite{Hall:2009bx}. Once the FI production ceases both sectors evolve independently. Entropy conservation in each sector separately can be then used to obtain the energy density evolution, provided that there are no other sources of heat exchange and the dark sector is in equilibrium with itself.
Conversely, during the freeze-in stage the dynamical evolution of SIDM provides a rich interplay between the energy available within the dark sector and its capacity to transform it into number density through self-interactions while cooling the system~\cite{Bernal:2015xba,Heeba:2018wtf,March-Russell:2020nun,Bernal:2020gzm}. Accurate treatment of such boosting of the freeze-in process through self-thermalization requires a proper implementation of the temperature (or energy density) and number density evolution equations. In this work we aim to address this issue numerically without any constraining approximations, and thus differing from the existing literature, by solving the set of coupled Boltzmann equations (cBE) in the hydrodynamic approach~\cite{Binder:2017rgn,Binder:2021bmg}.

Noteworthy, a completely secluded dark sector does not predict any direct experimental signals due to the absence of non-gravitational interactions with regular matter. While, if the two sectors are \textit{almost} secluded, i.e. there exists a very weak portal between them, the potential detection while not hopeless, is still rather challenging. Indeed, current direct detection technology allows for testing of only certain scenarios involving feeble portals, provided the dark sector particle is not heavier than the MeV scale~\cite{Hambye:2018dpi,Bhattiprolu:2023akk}. Motivated by this, in this work we are interested in exploring the question whether or not SIDM realization can be made more predictive through existence of such portals. In particular, if it may lead to detectable signatures and can SIDM be produced solely via freeze-in, without any need for additional mechanisms. We address that question on a quantitative level in three simple scenarios extending the Standard Model (SM) by: a real scalar with broken $\mathbb{Z}_2$, a complex scalar with unbroken $\mathbb{Z}_3$, and a $\mathbb{Z}_3$ scalar with an additional scalar mediator. We aim to demonstrate that interactions within the dark sector can significantly modify the standard production of dark matter via the freeze-in mechanism, thereby affecting its experimental predictability.

The article is organized hierarchically, based on the complexity of the models that we will introduce henceforth, as follows. In Section 2 we briefly describe the set of cBEs in the momentum moments approach. In Section 3 we examine the simplest realization: a dark sector consisting of a singlet real scalar DM candidate with a broken $\mathbb{Z}_2$ symmetry, allowing DM to mix with the Higgs, and therefore decay. To avoid rapid decays, interactions with visible matter should be significantly suppressed, leading to the dark sector being naturally populated via the FI mechanism. Remarkably, this model differs from the SIMP scenario~\cite{Hochberg:2014dra} as it considers a colder dark sector than the SM, and from \cite{Hufnagel:2022aiz} by reintroducing the portal to matter, and from both by solving the system of cBE using the hydrodynamical approach. Unsurprisingly, this model is strongly constrained by the INTEGRAL and NuSTAR telescope observations~\cite{Laha:2020ivk}. Moreover, the broken phase further constraints the parameter space, as domain walls may arise during the dark phase transition~\cite{Saikawa:2017hiv}. These findings suggest that the freeze-in mechanism cannot successfully populate the dark sector in this model in its entirety, but an additional production mode is necessary.

Next, in Section 4 we study a stable complex scalar DM candidate carrying a charge under an unbroken $\mathbb{Z}_3$ symmetry.
This model shares similar dynamical features with its predecessor, with however, one crucial distinction: here the DM is stable. Its interactions with matter are mediated by the Higgs boson, and are inherently suppressed, naturally out of the reach of direct or indirect detection experimental prospects. To generalize this model, in Section 5 we introduce a simple extension by a Higgs-like singlet real scalar mediator. Both the DM and the mediator are produced via the FI mechanisms and we solve a set of four cBEs accounting for the two particles (one for number density and one for temperature each). This approach considers all interactions between DM and the mediator, including annihilation-production and heat exchange processes, differing from standard literature where kinetic equilibrium within the dark sector is typically assumed. To the best of our knowledge such scenario has not been addressed in any other related work. In Section 6 we provide our conclusions.

\section{The Boltzmann equation and moments approach}
\label{sec:BoltzmannE}

In the homogeneous and isotropic universe expanding with the Hubble rate $H$ one can describe the evolution of a phase space distribution function $f_i(p,t)$ of a particle species $i$ through the Boltzmann equation (fBE):
\begin{equation}\label{eq:fBE}
  \frac{\partial f_i}{\partial t} - H p\frac{\partial f_i}{\partial p} = C[f_i]\,,
\end{equation}
where the right hand side is the collision operator encoding all the possible interactions between particle $i$ with itself and other states in the plasma. There has been a growing interest in the literature in approaching the determination of thermal production by directly analyzing the (numerical) solution of this equation  in order to obtain not only the number density of DM particles, but also their velocity distribution (see e.g.~\cite{Binder:2017rgn,Ala-Mattinen:2019mpa,Binder:2021bmg,Ala-Mattinen:2022nuj,Aboubrahim:2023yag,Kim:2023bxy,Beauchesne:2024zsq}). Moreover, it has been shown that in situations where an efficient equilibriation process is absent such a treatment can actually be necessary for achieving accuracy matching the observations~\cite{Binder:2017rgn,Hryczuk:2022gay}. 

Nevertheless, in models where substantial self-scatterings redistribute DM momenta efficiently and thus enforce the thermal \textit{shape}, albeit with potentially different normalization and temperature, a hydrodynamic approach that leads to a set of fluid equations is usually sufficient.\footnote{In fact, it can even provide a better estimate of the final relic abundance than the solution of Eq.~\eqref{eq:fBE}, if in the latter one does not incorporate the notoriously CPU expensive self-scatterings~\cite{Hryczuk:2022gay}.} In that case, instead of solving the full Eq.~\eqref{eq:fBE}, one can consider its momentum moments, where the lowest one governs the evolution of the number density $n = g_i \int d^3p/(2\pi)^3\,f_i$, with $g_i$ denoting the number of internal degrees of freedom. Going one level up in the Boltzmann hierarchy and paremetrizing the second moment via the velocity dispersion and defining the \textit{temperature parameter}\footnote{Note, that this parameter coincides with the actual temperature for an equilibrium distribution in the Maxwell-Boltzmann (MB) approximation. For non-thermal distributions one can view it as an alternative definition of the DM ‘temperature’ valid in the dilute limit of the system. Such a limit is assumed throughout this work and is justified since the DM component during the freeze-in stage is very underabundant, while at later stages of the evolution becomes non-relativistic.}, $T_i =g_i/(3n)\int d^3p(2\pi)^{-3}(p^2/E)f_i$, one arrives at a set of \textit{coupled Boltzmann equations} (cBE)~\cite{Bringmann:2006mu,Binder:2021bmg}
\begin{equation}\label{eq:cBE}
\begin{split}
    \frac{Y'}{Y} &= \frac{1}{x\tilde H}\braket{C}\,,\\
    \frac{x_\text{ds}'}{x_\text{ds}} &=-\frac{1}{x \tilde H} \braket{C}_2  
    +\frac{Y'}{Y} - \frac{H}{x \tilde H}\frac{\braket{p^4/E^3}}{3T_i}-\frac{2s'}{3s}\,,
\end{split}
\end{equation}
where $s$ is the SM entropy, $Y = n/s$, $x_\text{ds}=m_i/T_i$, $\tilde H=H(T)/(1+3T(dg_\text{eff}^s/dT)/g_\text{eff}^s)$, $'= d/dx$ and $x = m_i/T$. The thermal averages are defined as $\braket{\mathcal{O}}=g_i n^{-1}(2\pi)^{-3}\int d^3p\,\mathcal{O}f_i$ and $\braket{\mathcal{O} }_2 = g_i(3T_i n)^{-1}(2\pi)^{-3}\int d^3p\,(p^2/E)\,\mathcal{O}f_i$. This form of cBE is obtained through closing the Boltzmann hierarchy by assuming the distribution with equilibrium shape within the MB approximation
\begin{equation}
    f_i(p)=e^{\mu_i/T_i} e^{-E/T_i} = (n/n^\text{eq})f_i^{\text{eq}}(p;T_i).
\end{equation}
In the following we adopt this hydrodynamical approach since the existence of a cannibal phase is inherently linked to strong $2\to2$ self-scatterings.\footnote{By construction the $\sigma_{2\to 3}$ cross section is assumed to be large enough to lead to an efficient $2\to 3$ process and therefore $2\to2$ scatterings are expected to be even more frequent, as the corresponding  $\sigma_{2\to 2}$ is lower order in the coupling constant.}
For every of the discussed models we provide below the corresponding set of cBE with the explicit form of both moments of the collision term.

\section{The simplest case: SM $+$ a real scalar}
\label{sec:Z2}

In this section we discuss the cannibal phase of the arguably simplest possible dark matter model, i.e., a theory obtained by adding to the Standard Model only one real scalar field describing the DM. A completely secluded realization of this scenario has been studied recently in \cite{Hufnagel:2022aiz} where it was found that it is indeed still a viable possibility, although rather difficult to test experimentally. We start our analysis from the very same model, augmented with a potentially non-zero Higgs portal (HP) coupling, to first discuss all the relevant processes and showcase the formalism, and second to answer the question of whether including the portal interaction may give rise to detectable signatures.

\subsection{The model}\label{sec:Z2_model}

The model consists of a singlet real scalar field $\varphi$ stabilized by a $\mathbb{Z}_2$ symmetry~\cite{Silveira:1985rk}. This field naturally couples to the Higgs doublet $H$ through the HP interaction, 
\begin{equation}\label{eq:HP}
V_\text{HP}(H,\varphi)=
\frac{1}{2}\lambda_{h\varphi}\varphi^2H^\dagger H\,,
\end{equation}
and its self-interactions are encoded in the potential
\begin{equation}\label{eq:Vself}
    V_\text{self}(\varphi) = \frac{1}{2!}\mu^2\varphi^2+\frac{\lambda}{4!}\varphi^4\,.
\end{equation}
When $\lambda_{h\varphi}\gtrsim \mathcal{O}(10^{-4})$  the interaction mediated by the Higgs can be strong enough for the thermal freeze-out to lead to the observed relic abundance. This region of the parameter space we refer to as the WIMP limit, because then $\varphi$ has similar phenomenological properties to any other massive particle with weak scale interactions. Such realization of the scalar singlet DM still remains viable, however, it is under tension with direct detection (DD) and indirect detection (ID) experiments (see e.g.~\cite{Casas:2017jjg,GAMBIT:2017gge} and references therein) with the remaining allowed mass ranges that provide correct relic density being a) very close to the Higgs resonance or b) at the TeV scale.

Alternatively, $\varphi$ can also be a Feebly Interacting Massive Particle (FIMP) candidate if $\lambda_{h\varphi}\lesssim \mathcal{O}(10^{-7})$. For such small couplings it does not attain thermal equilibrium in the early Universe and the main mechanisms for its production are freeze-in and gravitational~\cite{Lebedev:2021xey}. If produced mostly via freeze-in, the requirement of correct relic density sets the expected value of the coupling to be $\lambda_{h\varphi}\sim \mathcal{O}(10^{-9})$ for a mass in the GeV range~\cite{Lebedev:2019ton}.

The scenario we focus on here serves as yet another possibility: the $\mathbb{Z}_2$ symmetry is broken explicitly or spontaneously leading to cubic terms in the potential \eqref{eq:Vself} and therefore to cannibalizing $3\leftrightarrow2$ processes.\footnote{When the $\mathbb{Z}_2$ symmetry remains unbroken, the dominant cannibalizing process affecting the number density of $\varphi$ is the $4\leftrightarrow2$ self-number changing reaction, whose matrix element is suppressed by $\lambda^2$ at tree level~\cite{Arcadi:2019oxh}.} The downside of such realization is that it removes the symmetry protection for DM stability, making it inherently unstable. However, since aside gravitational the only interaction with the SM is through the HP, the $\varphi$ lifetime can be made extremely long if only the $\lambda_{h\varphi}$ coupling is small enough. To avoid this complication, the analysis in \cite{Hufnagel:2022aiz} assumed $\lambda_{h\varphi}=0$, resulting in a secluded and essentially undetectable dark sector. Instead, we relieve this assumption and study also the case with $\lambda_{h\varphi}\neq 0$ and specifically when the $\mathbb{Z}_2$ symmetry breaking is triggered by the $\varphi$ field obtaining a vacuum expectation value (VEV)
\begin{equation}\label{eq:vevphi}
   \braket{\varphi} =  \omega = \pm \sqrt{\frac{3}{\lambda}}\sqrt{v^2\lambda_{h\varphi} -2 \mu^2}\,,
\end{equation}
where $v$ is the SM VEV. The symmetry breaking leads to $\varphi$ mixing with the Higgs, which culminates in the emergence of a Higgs-like scalar~\cite{Falkowski:2015iwa} with a decay rate into SM states is proportional to $\sin^2\theta$ and accordingly the lifetime $\tau_\varphi\propto 1/\sin^2\theta$.\footnote{The rotation matrix parametrized with a rotation angle $\theta$ is given in the appendix~\ref{ap:vac_stab}.} Thus, indeed $\theta\ll 1$ becomes crucial to ensure that the lifetime is significantly larger than the age of the universe.\footnote{Indirect detection experiments impose even more stringent constraints on the lifetime of decaying dark particles. For instance, the INTEGRAL and NuSTAR experiments rule out lifetimes shorter than $\sim 10^{27}$~s for decaying DM with a mass of $\sim1\,\text{MeV}$ ~\cite{Laha:2020ivk} (the age of the universe is approximately $4\times 10^{17}\,\text{s}$).} This condition can be expressed in terms of $\lambda_{h\varphi}$ by noting that the scalar VEV is
\begin{equation}
        \\w=\sqrt{\frac{3}{\lambda}}\left(m_\varphi+\frac{3m_\varphi v^2}{\lambda(2m_h^2-2m_\varphi^2)}\lambda_{h\varphi}^2\right) + \mathcal{O}(\lambda_{h\varphi}^3)\,.
\end{equation}
This implies that the mixing angle is at leading order in $\lambda_{h\varphi}$ (and assuming $m_\varphi\ll m_h$)
\begin{equation}\label{eq:DM_Higgs_mixing}
    \theta\approx \frac{2\sqrt{3}\lambda_{h\varphi}m_\varphi v}{(m_\varphi^2-m_h^2)\sqrt{\lambda}}\implies \Gamma_{\varphi\to\text{SM}\,\text{SM}}^\text{tree}\propto \frac{\lambda_{h\varphi}^2m_\varphi^2v^2}{m_h^4\lambda}\,.
\end{equation}

Let us note in passing that the spontaneous $\mathbb{Z}_2$ symmetry breaking results in a dark phase transition that potentially leads to the formation of domain walls, as noted in~\cite{Saikawa:2017hiv}, which can dominate the energy density of the early universe. To prevent this, the surface tension $\sigma$ of the domain wall should be bounded from above by the MeV scale, which implies that $\sqrt{\lambda}w^3 \lesssim \text{MeV}$, given $\sqrt{\lambda}w^3\sim \sigma$~\cite{Zeldovich:1974uw}. On its face value, such a constraint effectively rules out the accessible parameter space of this specific scenario. These are, however, reliant of assumptions that need not hold, and there are cases where these bounds do not apply at all.\footnote{For instance, a reheating temperature of the dark sector lower than the temperature of dark phase transition or unstable domain walls~\cite{Hufnagel:2022aiz}. Another alternative involves a dark sector with an explicit breaking of the $\mathbb{Z}_2$ symmetry.}  Since the focus of this work is independent on how the issue of domain walls constraint is settled, we will not discuss this issue further. Especially, given that as we will see below, the limits on $\lambda_{h\varphi}$ in this particular model are stringent enough to render the impact of the HP on the DM production essentially negligible, prompting the analysis of more promising models in the next sections.

\subsection{The $\lambda_{h\varphi}=0$ case}\label{sec:Z2_lhphi_neq_0}

\begin{figure}[t!]
  \centering
  \begin{tabular}{c@{\hspace{1.2cm}}c@{\hspace{1.2cm}}c}
    \begin{tikzpicture}[baseline=(b)]
      \begin{feynman}
        \vertex (a1);
        \vertex [below right=1.1cm of a1] (b);
        \vertex [below=0.7cm of b](a2);
        \vertex [right=1.5cm of b](c);
        \vertex [above right=1.2cm of c](d1);
        \vertex [below right=1.2cm of c](d2);
        \vertex [above left=1.1cm of a2](e1);
        \vertex [below left=1.1cm of a2](e2);
        \diagram* {
          (a1) -- [scalar] (b),
          (a2) -- [scalar] (b) --[scalar](c),
          (c) -- [scalar](d1),
          (c) -- [scalar](d2),
          (a2) -- [scalar](e1),
          (a2) -- [scalar](e2),
        };
      \end{feynman}
    \end{tikzpicture} &

    \begin{tikzpicture}[baseline=(in)]
      \begin{feynman}
        \vertex [dot] (a);
        \vertex [below =0.7cm of a] (in);
        \vertex [below =1cm of in] (b);
        \vertex [left =0.7cm of a] (i1);
        \vertex [left =1.5cm of b] (i2) ;
        \vertex [right =1.5cm of b] (f1) ;
        \vertex [right =1.5cm of a] (f2) ;
        \vertex [below left=1cm of i1] (ii2) ;
        \vertex [left=0.8cm of i1] (ii1) ;
        \diagram* {
          (i1) -- [scalar](a)
            -- [scalar] (f2),
          (a) -- [scalar] (in),
          (in) -- [scalar] (b),
          (i2) -- [scalar] (b),
          (b) -- [scalar] (f1),
          (ii1) -- [scalar] (i1),
          (ii2) -- [scalar] (i1),
        };
      \end{feynman}
    \end{tikzpicture} &

    \begin{tikzpicture}[baseline=(in)]
      \begin{feynman}
        \vertex [dot] (a);
        \vertex [below =0.8cm of a] (in);
        \vertex [below =0.8cm of in] (b);
        \vertex [left =1.5cm of in] (i3);
        \vertex [left =1.5cm of a] (i1);
        \vertex [left =1.5cm of b] (i2);
        \vertex [right =3.2cm of i2] (f1);
        \vertex [right =3.2cm of i1] (f2);
        \diagram* {
          (i1) -- [scalar](a)
            -- [scalar] (f2),
          (a) -- [scalar] (b),
          (i2) -- [scalar] (b),
          (b) -- [scalar] (f1),
          (i3) -- [scalar] (in),
        };
      \end{feynman}
    \end{tikzpicture} \\ \\ 

    \begin{tikzpicture}[baseline=(g)]
      \begin{feynman}
        \vertex (a1);
        \vertex [below right=1.2cm of a1] (b);
        \vertex [left =1.2cm of b] (g);
        \vertex [below left=1.2cm of b](a2);
        \vertex [right=of b](c);
        \vertex [above right=1.1cm of c](d1);
        \vertex [below right=1.1cm of c](d2);
        \diagram* {
          (a1) -- [scalar] (b),
        (a2) -- [scalar] (b) -- [scalar] (c),
        (c) -- [scalar](d1),
        (c) -- [scalar](d2),
        (g) -- [scalar](b),
        };
      \end{feynman}
    \end{tikzpicture} &

    \begin{tikzpicture}[baseline=(g)]
      \begin{feynman}
        \vertex (a1);
        \vertex [below right=1.1cm of a1] (b);
        \vertex [left =1.1cm of b] (g);
        \vertex [below left=1.1cm of b](a2);
        \vertex [right=of b](c);
        \vertex [above right=1.1cm of c](d1);
        \vertex [below right=1.1cm of c](d2);
        \diagram* {
          (a1) -- [scalar] (c),
          (a2) -- [scalar] (b) -- [scalar] (c),
          (c) -- [scalar](d1),
          (c) -- [scalar](d2),
          (g) -- [scalar](b),
        };
      \end{feynman}
    \end{tikzpicture} &

    \begin{tikzpicture}[baseline=(in)]
      \begin{feynman}
        \vertex [dot] (a);
        \vertex [below=0.6cm of a] (in);
        \vertex [below=1cm of in] (b);
        \vertex [left =1.5cm of a] (i1);
        \vertex [below=1cm of i1] (ii2);
        \vertex [left =1.5cm of b] (i2);
        \vertex [right=1.5cm of b] (f1);
        \vertex [right=1.5cm of a] (f2);
        \diagram* {
          (i1) -- [scalar](a)
            -- [scalar] (f2),
          (a) -- [scalar] (b),
          (i2) -- [scalar] (b),
          (b) -- [scalar] (f1),
          (ii2) -- [scalar] (a),
        };
      \end{feynman}
    \end{tikzpicture}
  \end{tabular}
  \caption{Feynman diagrams for the $3\leftrightarrow2$ reaction. In the limit $\lambda_{h\varphi}\to 0$ all initial, virtual and final states are $\varphi$.}
  \label{fig:feynman3-2}
\end{figure}
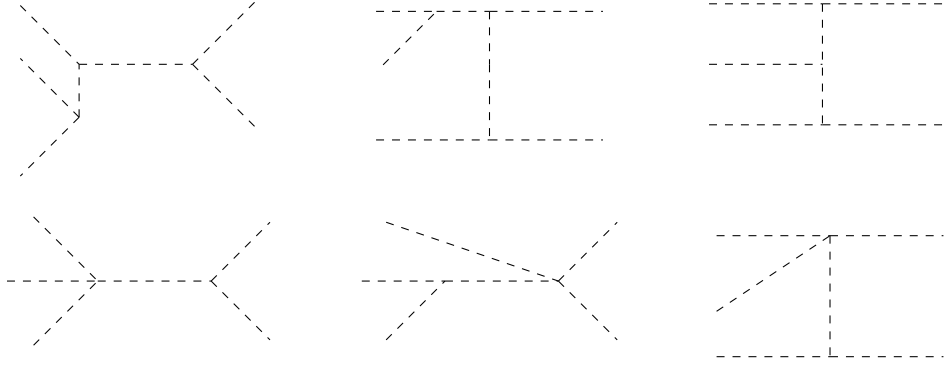

After spontaneous symmetry breaking, the potential given by Eq.~\eqref{eq:Vself} takes the form
\begin{equation}\label{eq:Z2_broken_phase_potential}
    V_\text{self} = \frac{1}{2}m_\varphi^2\varphi^2+\frac{g}{3!}\varphi^3+\frac{\lambda}{4!}\varphi^4\,,
\end{equation}
where the coupling of the cubic term is related to other parameters via $g = \sqrt{3\lambda}m_\varphi$ and the physical squared mass is $m_\varphi^2 = 2\vert \mu\vert^2 = \lambda v^2/3$ \cite{Hufnagel:2022aiz}. The primary contribution at lowest order in $\lambda$ corresponds to the $3\varphi\leftrightarrow 2\varphi$ self-number changing reaction, whose matrix element is presented in Eq.~\eqref{eq:mat_el_real}. The tree level Feynman diagrams involved in the reaction are shown in Figure~\ref{fig:feynman3-2}.

In this scenario the absence of a connection between the both sectors forbids the exchange of entropy, allowing them to evolve independently. Therefore, the standard assumption of initial kinetic equilibrium ($T_\varphi^i=T^i$, with $T$ the temperature of the SM) is not justified. This renders an extra degree of freedom in the parameter space, i.e. an initial condition after reheating, here represented via the initial ratio of temperatures $\xi_\infty=T^i_\varphi/T^i$. In fact, this observation is crucial for the viability of the model, since number changing self-interactions in effect heat the dark sector during the dark freeze-out. This renders a conflict between obtaining correct abundance and predicting successful structure formation \cite{1992ApJ...398...43C,deLaix:1995vi,Chatterjee:2019olf,Irsic:2017ixq}, unless the dark sector is significantly colder than the SM or has an efficient way of dissipating the excess heat into the SM plasma.

The set of equations that determine the evolution of the system is provided by Eq.~\eqref{eq:cBE}. The collision operator is given by
\begin{equation}
\begin{split}
    C_{3\varphi\leftrightarrow 2\varphi}=\frac{1}{2E_\varphi g_\varphi}\int  \Bigg(&- f_\varphi(p)\vert\tilde{\mathcal{M}}_{\underline{\varphi}2\to 345}\vert^2 f_2\,d\Pi_2 \left(\frac{1}{3!}d\tilde \Pi_3 d\tilde\Pi_4 d\tilde\Pi_5\,\right)
     \\&+(1\! +\! f_\varphi(p))\vert\tilde{\mathcal{M}}_{12\to \underline{\varphi}45}\vert^2\left(\frac{1}{2!}d\Pi_1 d\Pi_2\, f_1 f_2 \right)
     \left(\frac{1}{2!} d\tilde \Pi_4 d\tilde\Pi_5 \right) 
    \\& -f_\varphi(p)\vert\tilde{\mathcal{M}}_{12\leftarrow \underline{\varphi}45}\vert^2\left(\frac{1}{2!}d\Pi_4 d\Pi_5\, f_4 f_5 \right)\left(\frac{1}{2!} 
    d\tilde\Pi_1 d\tilde\Pi_2\right)
   \\& +(1+f_\varphi(p))\vert\tilde{\mathcal{M}}_{\underline{\varphi}2\leftarrow 345}\vert^2
    \left(\frac{1}{3!} d\Pi_3 d\Pi_4 d\Pi_5\,f_3 f_4 f_5\right)
    d\tilde\Pi_2\Bigg)\,,
\end{split}
\end{equation}
where we label the momenta as $12\leftrightarrow 345$, with $\vert\tilde{\mathcal{M}}\vert^2=(2\pi)^4\delta^{(4)}\left(\sum_f p_f - \sum_i p_i\right)\vert\mathcal{M}_{3\leftrightarrow 2}\vert^2$, and define $d\tilde\Pi_i = d\Pi_i(1+f_i)$. The notation underlines the production and annihilation of the $\varphi$ states, also highlighting the symmetry factors. The Bose-Einstein enhancement factors can be safely neglected, $(1+f)\approx 1$, as the system is either diluted or non-relativistic, in the early and later stages of the evolution, respectively. 
The zeroth and second moment terms are detailed in Eq.~\eqref{eq:C0_self} and Eq.~\eqref{eq:C2_self}, respectively.

Previous work~\cite[Figure 3]{Hufnagel:2022aiz} demonstrates that the available parameter space for the model resides within the sub-GeV dark matter mass range. This preference arises because self-number changing reactions scale inversely with the fifth power of the DM mass ($\braket{\sigma_{3\to 2}v^2}\sim\lambda^3 m_\varphi^{-5}$, cf. Eq.~\eqref{eq:ap_m_scaling}). As a result, larger DM masses necessitate a stronger self-interaction coupling, which can violate perturbativity or unitarity constraints. Therefore, for this model we focus on sub-GeV dark matter masses.

\subsection{The $\lambda_{h\varphi}\neq 0$ case and relativistic freeze-in}\label{sec:Z2_lhphi_neq}

A secluded cannibal dark sector remains agnostic about the mode of its initial production. That is, the initial $\xi_\infty$ (or initial abundance and temperature, or even more generally $f(p)$) are presumed. The introduction of a HP interaction opens a possibility of populating the dark sector solely by the freeze-in mechanism. Moreover, if the portal is substantial, it may also facilitate heat exchange which is a necessary ingredient of the mechanism ascribed as the SIMP miracle~\cite{Hochberg:2014dra,Hochberg:2014kqa}.

When the number changing self-interactions are absent, it is usually sufficient to solve Eq.~\eqref{eq:fBE} at the lowest moment (i.e., the number density evolution) during freeze-in, as the transfer of heat does not impact the final DM abundance.\footnote{Although, solving the cBE (or fBE) may be essential to determine if the amount of energy injected into the DM fluid by the freeze-in mechanism conflicts with Lyman-$\alpha$ forest data, which constrains the free-streaming length to $\lambda_\text{FS}<0.24\,\text{Mpc}$~\cite{Garzilli:2019qki}. } In our scenario, however, the dark sector can achieve chemical equilibrium through $2\leftrightarrow 3$ processes. While doing so, it can convert its kinetic energy into number density, a mechanism intriguingly opposite to the cannibalization phase~\cite{March-Russell:2020nun, Bernal:2015xba, Heeba:2018wtf}. Thus, it becomes crucial to solve the system of cBE, Eq.~\eqref{eq:cBE}, which takes the form
\begin{equation}
    \begin{split}
        \frac{Y_\varphi'}{Y_\varphi} &\supset +\frac{1}{x\tilde H}\left(\braket{C_{h\to \varphi\varphi}}
        +\braket{C_{hh\to \varphi\varphi}}\right)\Theta(T-T_\text{EWPT})\,,
        \\
        -\frac{x_\varphi'}{x_\varphi}&\supset +\frac{1}{x\tilde H}\left(\braket{C_{h\to  \varphi\varphi}}_ 2+\braket{C_{hh\to \varphi\varphi}}_2\right) \Theta(T-T_\text{EWPT}) \,.
    \end{split}
\end{equation}
As an initial condition we take $n_\varphi^i = n_\varphi^\text{eq}(T_\varphi^i)$, with $\xi_\infty=T_\varphi^i/T^i<1$ being a free parameter and $T_i=T_\text{EWPT}=150\,\text{GeV}$, where $T_\text{EWPT}$ is the temperature of the SM plasma at the Electroweak Phase Transition (EWPT).

Two comments are in order. First, this set of cBE focuses on production \textit{after} EWPT, neglecting contributions from \textit{before} and \textit{during} EWPT. Given our assumption of sub-GeV DM and therefore the limit $m_\varphi \ll m_h$, the main production is driven by Higgs decay, $h \to \varphi\varphi$, at a time when $T\sim 40\,\text{GeV}$~\cite{Heeba:2018wtf}.
During EWPT, the physical mass of the Higgs boson vanishes for a short period of time and then increases with decreasing temperature. Therefore, there is a point in time where $m_h(T)\approx m_\varphi$ and the mixing angle~\eqref{eq:DM_Higgs_mixing} is enhanced, thus DM can be produced through Higgs oscillations. This yield from this mode is approximately~\cite{Heeba:2018wtf}
\begin{equation}
    Y_{\varphi}^\text{EWPT} = \left(1.93\times 10^5\,\text{GeV}^{-4}\right)\,\lambda_{h\varphi}^2 m_\varphi^2 w^2 = 1.93\times 10^5\,\text{GeV}^{-4}\,\lambda_{h\varphi}^2\, \frac{3m_\varphi^4}{\lambda}\,,
\end{equation}
where we used Eq.~\eqref{eq:vevphi}. For $m_\phi =100\,\text{MeV}$ and $\lambda=10^{-2}$ this results in  $Y_\varphi^\text{EWPT} \approx 5.79\times 10^{-15}$, which is a small contribution compared to the one post-EWPT, (cf. Figure~\ref{fig:BM1_Z2}). 
Second comment is that the production from $h \to \varphi\varphi$ is essentially insensitive to the dynamics of the EWPT, and in fact also the exact value of $T_\text{EWPT}$, allowing us to adopt a simple Heaviside step function.

For the production from the Higgs decay the zeroth moment thermal average takes the analytical form
\begin{equation}
\begin{split}
\braket{C_{h\to \varphi\varphi}}=\frac{1}{n_\varphi} \frac{\lambda_{h\varphi}^2 v^2 m_h}{16\pi^3}\sqrt{1-\frac{4m_\varphi^2}{m_h^2}}T\,K_{1}(m_h/T)
\end{split}
\end{equation}
and for the second moment:
\begin{equation}\label{eq:FI_2}
    \braket{C_{h\to \varphi\varphi}}_2\approx\frac{1}{3n_\varphi T_\varphi} \frac{\lambda_{h\varphi}^2 v^2 m_h^2}{32\pi^3}\sqrt{1-\frac{4m_\varphi^2}{m_h^2}} T\,K_2(m_h/T)
\end{equation}
which is valid in the $m_\varphi \ll m_h$ limit which we assume in the numerical implementation (cf. Eq.~\eqref{eq:ap_Higgs_decay_C2}) and where $K_n$ are the modified Bessel functions of the second kind and order $n$. 
The sub-leading annihilation contributions are given in appendix \ref{sec:Higgs_ann}.

The interplay between freeze-in and number changing self-interactions process can lead to interesting dynamics. One can classify possible scenarios by the strength of the $2\leftrightarrow 3$ process:
\begin{itemize}
    \item Throughout the entire history of the dark sector, number changing reactions remain inefficient (i.e., $\Gamma_{2\leftrightarrow 3}\ll H$); this corresponds to the standard freeze-in mechanism. 
    \item The number changing self-interactions are inefficient initially, but become stronger in later stages of the evolution, e.g., following the EWPT. At that point the system strives to reach chemical equilibrium. During this phase $2\to 3$ reactions rapidly produce more DM while at the same time cooling the dark sector~\cite{Bernal:2020gzm,March-Russell:2020nun}.
    \item The number changing self-interactions remain efficient throughout the entire evolution of the system. As a result, FI production is predominantly supported through the process $2\to 3$. This differs from the previous point, as the dark sector is \textit{always} in chemical equilibrium with itself and there are no sudden attempts to re-establish equilibrium.
\end{itemize}

\subsection{Results}\label{sec:Z2_results}

\begin{figure}[t!]
    \centering
    \includegraphics[width=0.7\textwidth]{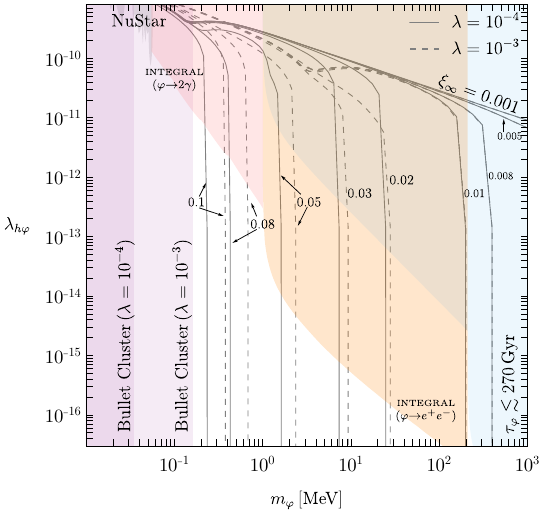}
    \caption{Limits on the parameter plane $\lambda_{h\varphi}\,\text{vs}\ m_\varphi$ for the scalar singlet DM model with spontaneously broken $\mathbb{Z}_2$ for two different values of the self-coupling $\lambda$. The gray solid and dashed lines shows contours giving the observed relic abundance for several values of $\xi_\infty$. The constraints are divided into three regions. The first region (purple), $m_\varphi\lesssim 100\,\text{keV}$, is in conflict with the Bullet Cluster observations~\cite{Randall:2008ppe}. The second region (gray, pink and orange), $10\,\text{keV}\lesssim m_\varphi<2m_\mu$, faces stringent constraints from decaying DM into $2\gamma/\,e^+e^-$ signals by the INTEGRAL and NuStar telescopes~\cite{Laha:2020ivk,Slatyer:2016qyl}. The third region (blue), $m_\varphi>2m_e$, partially overlaps the previous region and results in decay into $e^+e^-$ pairs (correspondingly $\mu^+\mu^-$ pairs when kinematically allowed) with lifetimes significantly shorter than those that are constrained by CMB data, $\tau_\varphi\lesssim 270\,\text{Gyr}$~\cite[Table 2]{Nygaard:2020sow}. 
    }
    \label{fig:resultsZ2}
\end{figure}
As was already mentioned, once $\lambda_{h\varphi} \neq 0$ this particular realization of the scalar singlet DM with spontaneously broken $\mathbb{Z}_2$ is strongly constrained by the stability requirement. Here we first quantify this statement and then answer the question of whether including such portal coupling can lead to detectable signals. Finally, we use this model as an illustration of the possible interesting dynamics encoded in the interplay of the processes discussed above.

In Figure~\ref{fig:resultsZ2} the allowed parameter space in the plane of $\lambda_{h\varphi}$ vs
$m_\varphi$ is presented. If $m_\varphi$ is above the threshold of $\mu^+\mu^-$ decay, then only extremely small values of $\lambda_{h\varphi}$ are not excluded by the DM stability condition ($\tau_\varphi\lesssim\,270\,\text{Gyr}$~\cite[Table 2]{Nygaard:2020sow}). For $2m_e<m_\varphi<2m_\mu$ the lifetime is most strongly constrained by INTEGRAL data with condition $\tau_\varphi\gtrsim 10^{27}\,\text{s}$~\cite{Laha:2020ivk}. Below the $e^+ e^-$ threshold the limits get weaker, but still quite significant, as only direct annihilation to photons is possible. 

Another potential conflict with observations may arise from disrupting the Big Bang Nucleosynthesis (BBN, $T\sim 0.1-1\,\text{MeV}$). However, this requires much shorter lifetimes,  $\tau_\varphi\lesssim 10^{12}\,\text{s}$~\cite{Depta:2020zbh}, than the ones already excluded by INTEGRAL or NuStar observations. Therefore such BBN constraints are subdominant and we choose to omit them in Figure~\ref{fig:resultsZ2}.

Finally, self-interactions of DM are constrained by observational data of DM elastic self-scattering in the galaxies and galaxy clusters. In particular, we show the exclusion limit from the Merging Galaxy Cluster 1E 0657-5 (the Bullet Cluster)~\cite{Randall:2008ppe}, $\sigma_T/m_\varphi<1\,\text{cm}^2/\text{g}$ at a typical velocity of $v=10^{-4}$, where the transfer cross section is defined as
\begin{equation}\label{eq:sigmaT}
    \sigma_T=\int d\Omega\,(1-\cos\alpha)\frac{d\sigma_{2\to 2}}{d\Omega}\,,
\end{equation}
with $\sigma_{2\to 2}$ accounting for self-scattering in the form $\varphi\varphi \to \varphi\varphi$. For the scalar singlet model this constraint takes a particularly simple form of the allowed region satisfying $m_\varphi/16.32\,\text{MeV}\gtrsim \lambda^{2/3}$ and in contrast to the limits coming from DM stability it depends on the value of the quartic coupling $\lambda$. In Figure~\ref{fig:resultsZ2} we chose two representative values: $\lambda=10^{-4}$ and $\lambda=10^{-3}$, for which we also show as gray lines (solid and dashed, respectively) the contours for the parameters that lead to the relic density matching the observed value for different assumptions regarding $\xi_\infty$. The larger $\xi_\infty$ is, the larger the initial population and thus smaller additional production from FI is required; hence, smaller values of $\lambda_{h\varphi}$. The shape of the contours clearly indicates, as one would expect, that for small enough $\lambda_{h\varphi}$, the impact of the HP becomes negligible. 

The choice of $\lambda$ impacts the results in three ways. First, for which values of $m_\varphi$ the process $2\to 3$ dominates over $3\to 2$ throughout the dynamical evolution of the system. This manifests around $m_\varphi \simeq 200\,\text{keV}$ ($m_\varphi = 2\,\text{MeV}$) for $\lambda=10^{-4}$ ($\lambda=10^{-3}$) in Figure~\ref{fig:resultsZ2}, where solutions exhibit smaller values of $\lambda_{h\varphi}$ along the solution lines. Second, increasing $\lambda$ will lead to more DM depletion via the cannibal process, which can be compensated by increased production during FI, thereby necessitating a higher $\lambda_{h\varphi}$. And finally, on the self-scattering constraints, by shifting the excluded region it to the right (left) if $\lambda$ increases (decreases).

All in all, these results strongly suggest that for this particular model \AH{a} rather substantial initial population, coming from an additional production mechanism, is required to allow for long enough lifetime to be consistent with observations and therefore that freeze-in alone is insufficient to populate the dark sector. Additionally, in the parameter regions where the model is still allowed the value of $\lambda_{h\varphi}$ needs to be small enough that has a very weak to negligible effect on the DM production -- both through freeze-in process and through the heat transfer due to elastic scatterings.
\begin{figure}[t!]
    \centering
    \includegraphics[width=0.47\textwidth]{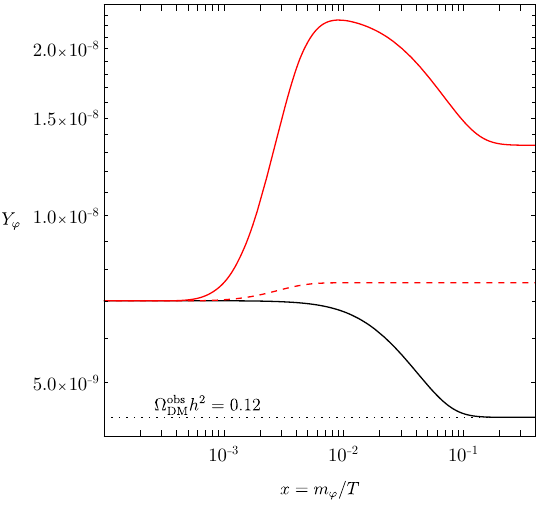}
    \hfill
    \centering
    \includegraphics[width=0.45\textwidth]{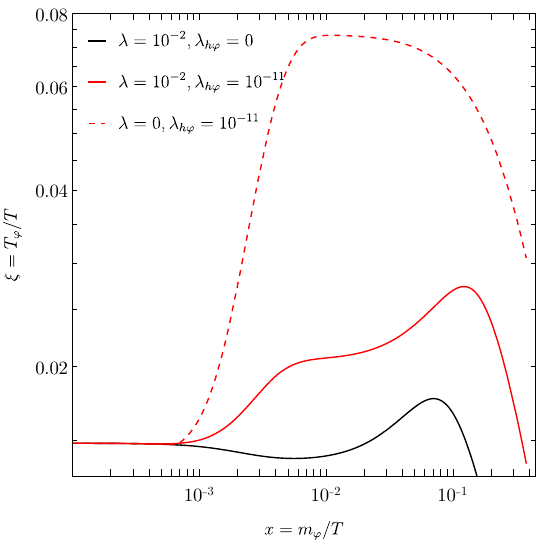}
    \caption{Evolution of the system for an example point: $m_\varphi = 100\,\text{MeV}$ and $\xi_\infty = 0.014$. The case $\lambda_{h\varphi} = 0$ matches the observed relic abundance (black line). When $\lambda_{h\varphi}= 10^{-11}$ (red lines), the system receives injection of energy and $\varphi$ particles from the Higgs decay. In the absence of self-interactions (red dashed) the increase in the yield is moderate, in contrast to the scenario where self-interactions are efficient (red thick), where the dark sector converts this injected energy into an increased number density.}
    \label{fig:BM1_Z2}
\end{figure}

Irrespective of quite limited viability of this simple model, it can serve as a clear illustration of dynamics of the freeze-in dynamics coupled with $2\leftrightarrow 3$ processes. In~Figure~\ref{fig:BM1_Z2} the evolution of the DM yield and temperature for an example benchmark point is presented for three different choices of the coupling values. Black lines show the baseline case of $\lambda_{h\varphi}=0$, while red one with the Higgs portal switched on, $\lambda_{h\varphi}=10^{-11}$, leading to a freeze-in contribution, and two different values for the self-coupling $\lambda$ governing the cannibal process. Without the HP coupling the system undergoes only dark sector freeze-out with a pronounced self-heating period around $x\sim 10^{-1}$. Introducing non-zero $\lambda_{h\varphi}$ leads to an additional injection of $\varphi$ particles (mostly) from Higgs decay that not only increase the yield but also are much more energetic further heating the system. And then it is crucial whether the $2\leftrightarrow 3$ self-interactions are strong enough to convert these excess heat into more $\varphi$ particle (solid line) or not (dashed line). This large increase in the yield, with corresponding decrease in temperature, has been noted in \cite{Bernal:2020gzm} and dubbed ``boosting freeze-in through thermalization''. This effect underlines the necessity of careful evaluation of the temperature evolution alongside the number density, which we have done fully numerically by solving the cBE system with all the contributing processes.

\section{SM $+$ $\mathbb{Z}_3$ complex scalar}
\label{sec:Z3}

The limitation stemming from instability of DM if the stabilising $\mathbb{Z}_2$ symmetry is broken can be avoided without changing the particle content of the model by imposing a $\mathbb{Z}_3$ symmetry instead. This requires change from a real scalar $\varphi$ to a complex one, which we will call $S$. Such a $\mathbb{Z}_3$-stable scalar was first introduced in the context of neutrino physics~\cite{Ma:2007gq}. As a WIMP candidate, its phenomenology was initially addressed in~\cite{Belanger:2012zr} and further discussed in~\cite{Hektor:2019ote}. Here we focus on SIDM~\cite{Ghosh:2022asg} combined with the FIMP scenario~\cite{Choi:2016tkj}. Incidentally, $\mathbb{Z}_3$ symmetry allows for a cubic self-coupling leading naturally to cannibal-type reactions of the form $3\leftrightarrow 2$, making it a suitable candidate for a model with cannibal dark sector.

\subsection{The model}

To the SM we add a complex scalar $S$ charged under a hidden $\mathbb{Z}_3$ symmetry, $S\to e^{2q\pi i/3} S$, with $q=1$. The most general renormalizable potential is given by:
\begin{equation}\label{HP_Z3}
    V_\text{HP}(H,S)=V_s + \lambda_{hs}\vert H\vert^2\vert S\vert^2\,,
\end{equation}
where $V_s$ encodes self interactions of the scalar field $S$,
\begin{equation}\label{eq:VS}
    V_s =\mu_s^2\vert S\vert^2  + \frac{g_s}{3!}(S^3+(S^*)^3) + \frac{\lambda_s}{4}\vert S\vert^4\,.
\end{equation}
To ensure stability of the potential first we demand $\mu_s^2,\,\lambda_s>0$ and absorb the complex phase of the VEV, $v_s$, in the scalar field $S$.
We now obtain the extrema of the potential by solving $\left.\partial_S V_s\right\vert_{S,S^*=v_s} = \lambda_s v_s^3/2+g_s v_s^2/2 + \mu_s^2 v_s = 0$. There are three solutions,
\begin{equation}
    v_s^0=0\qquad\text{and}\qquad v_s^\pm = \frac{-g_s\pm\sqrt{g_s^2-8\lambda_s \mu_s^2}}{2\lambda_s}\,.
\end{equation}
If $g_s^2<8\lambda_s \mu_s^2$, then the only real solution is $v_s^0$. To express this constraint in a more convenient form, we introduce the dimensionless parameter $k=g_s^2/(3\lambda_s \mu_s^2)$,
with the stability constraint becoming $k<8/3$. Note that the physical mass is given by $m_s^2 = \lambda_s v_s^2 + \mu_s^2$ and corresponds to $\mu_s^2$ if the stability constraint is satisfied.

Due to charge conservation, the only allowed number-changing reactions within the dark sector are $SSS\leftrightarrow S^*S$ and $S^*S^*S\leftrightarrow SS$, along with their complex conjugates. The corresponding Feynman diagrams are shown in~\Cref{fig:feynman3-2_Z3,fig:feynman3-2_Z3_}. In order to solve the cBE for this realization, we have to specify the collision operator. Considering first the process $S^*S\leftrightarrow SSS$ and treating $S$ and $S^*$ as different sates, the collision operator for $S^*$ in its most general form can be expressed as
\begin{equation}
\begin{split}
 C_{S^*S\leftrightarrow SSS}[S^*]=\frac{1}{2E_{S^*}g_{S^*}}\int&\Bigg((1+f_{S^*})\vert\tilde{\mathcal{M}}_{345\to\underline{S^*}2}\vert^2\left( \frac{1}{3!}d\Pi_3d\Pi_4d\Pi_5f_3f_4f_5 \right)\,d\tilde\Pi_2
        \\
        &-f_{S^*} \vert\tilde{\mathcal{M}}_{\underline{S^*} 2\to 345}\vert^2 d\Pi_2f_2\left(\frac{1}{3!}d\tilde\Pi_3d\tilde\Pi_4d\tilde\Pi_5 \right)\Bigg)\,,
\end{split}
\end{equation}
whereas for $S^*S\leftrightarrow SSS$, the collision operator for $S$ is
\begin{equation}
    \begin{split}
        C_{S^*S\leftrightarrow SSS}[S] = \frac{1}{2E_{S}\,g_{S}}&\int\Bigg(-f_S\vert\tilde{\mathcal{M}}_{1\underline{S}\to 345}\vert^2d\Pi_1f_1\,\left(\frac{1}{3!}d\tilde\Pi_3d\tilde\Pi_4d\tilde\Pi_5\right)
        \\&+(1+f_S)\vert\tilde{\mathcal{M}}_{12\to \underline{S}45}\vert^2 \left(d\Pi_1d\Pi_2f_1f_2\right)\left(\frac{1}{2!}d\tilde\Pi_4d\tilde\Pi_5\right)
        \\&-f_S\vert\tilde{\mathcal{M}}_{\underline{S}45\to 12}\vert^2\left(\frac{1}{2!}d\Pi_4d\Pi_5f_4f_5\right)\left(d\tilde\Pi_1d\tilde\Pi_2\right)
        \\&+(1+f_S)\vert\tilde{\mathcal{M}}_{345\to 1\underline{S}}\vert^2\left(\frac{1}{3!}d\Pi_3d\Pi_4d\Pi_5f_3f_4f_5\right)d\tilde\Pi_1
        \Bigg)\,,
    \end{split}
\end{equation}
where, as in the previous model, $d\tilde \Pi_i=d\Pi_i(1+f_i)$. Assuming that the system is diluted or non-relativistic, $1+f_i\approx 1$.
The remaining collision operators, as well as its zeroth and second moments integrals can be found in Appendix~\ref{ap:C_Z3}. The FI production contribution is analogous to the previous section. We consider the two degrees of freedom of the complex scalar to account for DM, meaning we assume $n = n_{S} + n_{S^*}$ with $n_{S} = n_{S^*}$ (i.e. the dark sector has no initial asymmetry and there is no CP violation in the model). We also assume the same initial conditions as in the previous section.
\begin{figure}[t!]
  \centering
  \begin{tabular}{c@{\hspace{1.2cm}}c@{\hspace{1.2cm}}c@{\hspace{1.2cm}}}
    \begin{tikzpicture}[baseline=(in)]
   \begin{feynman}
        \vertex [dot] (a);  
        \vertex [below=0.6cm of a] (in);
        \vertex [below=1cm of in] (b);
        \vertex [left =1.5cm of a] (i1) {\(S\)};  
        \vertex [below=0.82cm of i1] (ii2){\(S\)};
        \vertex [left =1.5cm of b] (i2){\(S\)};
        \vertex [right=1.5cm of b] (f1){\(S^*\)};
        \vertex [right=1.5cm of a] (f2){\(S\)};
        \diagram* {
          (i1) -- [scalar](a)
            -- [scalar] (f2),
          (a) -- [scalar, edge label=\(S^*\)] (b),
          (i2) -- [scalar] (b),
          (b) -- [scalar] (f1),
          (ii2) -- [scalar] (a),
        };
      \end{feynman}
    \end{tikzpicture} &

    \begin{tikzpicture}[baseline=(in)]
      \begin{feynman}
        \vertex [dot] (a) ; 
        \vertex [below =1.6cm of a] (b);
        \vertex [left =0.85cm of a] (i1);  
        \vertex [left =1.5cm of b] (i2){\(S\)};
        \vertex [right =1.3cm of b] (f1){\(S^*\)};
        \vertex [right =1.3cm of a] (f2){\(S\)};  
        \vertex [below left=0.9cm of i1] (ii2) {\(S\)};
        \vertex [left=0.75cm of i1] (ii1) {\(S\)};
        \diagram* {
          (i1) --[scalar, edge label=\(S\)](a)
            -- [scalar] (f2),
          (a) -- [scalar,edge label=\(S\)] (b),
          (i2) -- [scalar] (b),
          (b) -- [scalar] (f1),
          (ii1) -- [scalar] (i1),
          (ii2) -- [scalar] (i1),
        };
      \end{feynman}
    \end{tikzpicture} &
    \begin{tikzpicture}[baseline=(g)]
      \begin{feynman}
        \vertex (a1) {\(S\)};  
        \vertex [below right=1.1cm of a1] (b);
        \vertex [left =1.1cm of b] (g){\(S\)};
        \vertex [below left=1.1cm of b](a2){\(S\)};
        \vertex [right=of b](c);
        \vertex [above right=1cm of c](d1){\(S\)};
        \vertex [below right=1cm of c](d2){\(S^*\)};
        \diagram* {
          (a1) -- [scalar] (c),
          (a2) -- [scalar] (b) -- [scalar,edge label'=\(S^*\)] (c),
          (c) -- [scalar](d1),
          (c) -- [scalar](d2),
          (g) -- [scalar](b),
        };
      \end{feynman}
    \end{tikzpicture}
  \end{tabular}
  \caption{Feynman diagrams for the $SSS\leftrightarrow S^*S$ reaction in the limit $\lambda_{hs}\to 0$.}
  \label{fig:feynman3-2_Z3}
\end{figure}
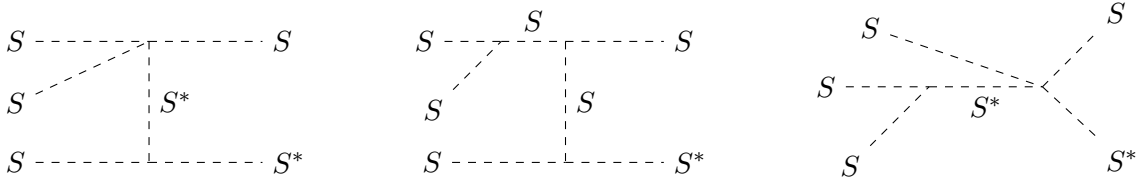
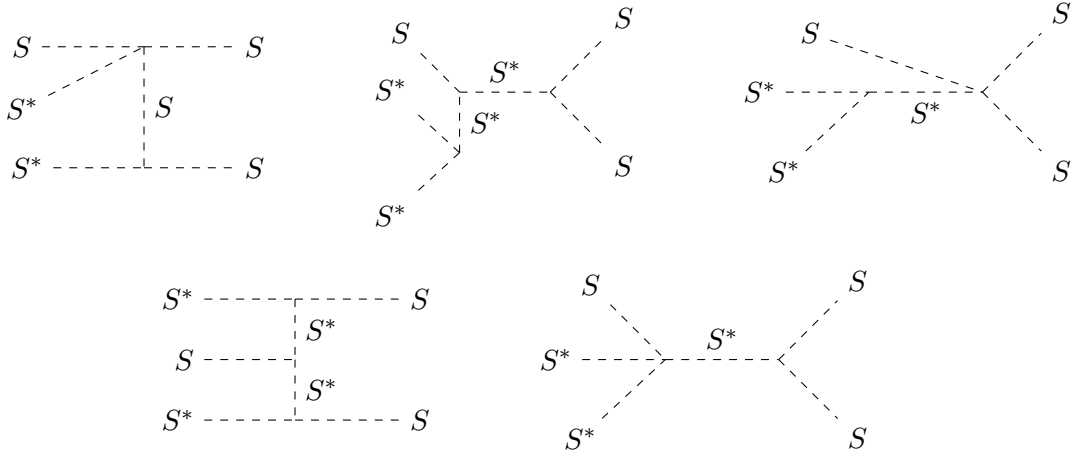
\begin{figure}[t!]
  \centering
  \begin{tabular}{c@{\hspace{1.2cm}}c@{\hspace{1.2cm}}c@{\hspace{1.2cm}}}
    \begin{tikzpicture}[baseline=(in)]
   \begin{feynman}
        \vertex [dot] (a);  
        \vertex [below=1cm of in] (b);
        \vertex [left =1.35cm of a] (i1) {\(S\)};  
        \vertex [below=0.82cm of i1] (ii2){\(S^*\)};
        \vertex [left =1.2cm of b] (i2){\(S^*\)};
        \vertex [right=1.2cm of b] (f1){\(S\)};
        \vertex [right=1.2cm of a] (f2){\(S\)};
        \diagram* {
          (i1) -- [scalar](a)
            -- [scalar] (f2),
          (a) -- [scalar, edge label=\(S\)] (b),
          (i2) -- [scalar] (b),
          (b) -- [scalar] (f1),
          (ii2) -- [scalar] (a),
        };
      \end{feynman}
    \end{tikzpicture} &

\begin{tikzpicture}[baseline=(b)]
      \begin{feynman}
        \vertex (a1){\(S\)};
        \vertex [below right=1.1cm of a1] (b);
        \vertex [below=0.8cm of b](a2);
        \vertex [right=1.2cm of b](c);
        \vertex [above right=1cm of c](d1){\(S\)};
        \vertex [below right=1cm of c](d2){\(S\)};
        \vertex [above left=0.8cm of a2](e1){\(S^*\)};
        \vertex [below left=0.8cm of a2](e2){\(S^*\)};
        \diagram* {
          (a1) -- [scalar] (b),
          (a2) -- [scalar, edge label'=\(S^*\)] (b) --[scalar, edge label=\(S^*\)](c),
          (c) -- [scalar](d1),
          (c) -- [scalar](d2),
          (a2) -- [scalar](e1),
          (a2) -- [scalar](e2),
        };
      \end{feynman}
    \end{tikzpicture} 
    &
    \begin{tikzpicture}[baseline=(g)]
      \begin{feynman}
        \vertex (a1) {\(S\)};  
        \vertex [below right=1.1cm of a1] (b);
        \vertex [left =1.1cm of b] (g){\(S^*\)};
        \vertex [below left=1.1cm of b](a2){\(S^*\)};
        \vertex [right=of b](c);
        \vertex [above right=1.1cm of c](d1){\(S\)};
        \vertex [below right=1.1cm of c](d2){\(S\)};
        \diagram* {
          (a1) -- [scalar] (c),
          (a2) -- [scalar] (b) -- [scalar,edge label'=\(S^*\)] (c),
          (c) -- [scalar](d1),
          (c) -- [scalar](d2),
          (g) -- [scalar](b),
        };
      \end{feynman}
    \end{tikzpicture}
    \end{tabular}
    \\[0.3cm]
  \centering
  \begin{tabular}{c@{\hspace{1.2cm}}c@{\hspace{1.2cm}}}
     \begin{tikzpicture}[baseline=(in)]
      \begin{feynman}
        \vertex  (a);
        \vertex [below =0.8cm of a] (in);
        \vertex [below =0.8cm of in] (b);
        \vertex [left =1.2cm of in] (i3){\(S\)};
        \vertex [left =1.2cm of a] (i1){\(S^*\)};
        \vertex [left =1.2cm of b] (i2){\(S^*\)};
        \vertex [right =3.2cm of i2] (f1){\(S\)};
        \vertex [right =3.2cm of i1] (f2){\(S\)};
        \diagram* {
          (i1) -- [scalar](a)
            -- [scalar] (f2),
          (a) -- [scalar,edge label=\(S^*\)] (in),
          (in) --[scalar,edge label=\(S^*\)] (b),
          (i2) -- [scalar] (b),
          (b) -- [scalar] (f1),
          (i3) -- [scalar] (in),
        };
      \end{feynman}
    \end{tikzpicture} &
\begin{tikzpicture}[baseline=(g)]
      \begin{feynman}
        \vertex (a1){\(S\)};
        \vertex [below right=1.4cm of a1] (b);
        \vertex [left =1.1cm of b] (g){\(S^*\)};
        \vertex [below left=1.1cm of b](a2){\(S^*\)};
        \vertex [right=of b](c);
        \vertex [above right=1.1cm of c](d1){\(S\)};
        \vertex [below right=1.1cm of c](d2){\(S\)};
        \diagram* {
          (a1) -- [scalar] (b),
        (a2) -- [scalar] (b) -- [scalar,edge label=\(S^*\)] (c),
        (c) -- [scalar](d1),
        (c) -- [scalar](d2),
        (g) -- [scalar](b),
        };
      \end{feynman}
    \end{tikzpicture}
  \end{tabular}
  \caption{Feynman diagrams for $S^*S^*S\leftrightarrow SS$ in the limit $\lambda_{hs}\to 0$.}
  \label{fig:feynman3-2_Z3_}
\end{figure}
%

\subsection{Results}\label{sec:Z3_results}

\begin{figure}[t!]
    \centering
    \includegraphics[width=0.49\textwidth]{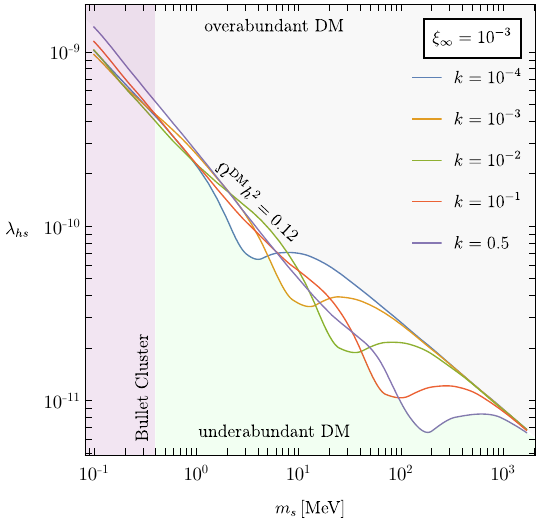}
    \hfill
    \centering
    \includegraphics[width=0.49\textwidth]{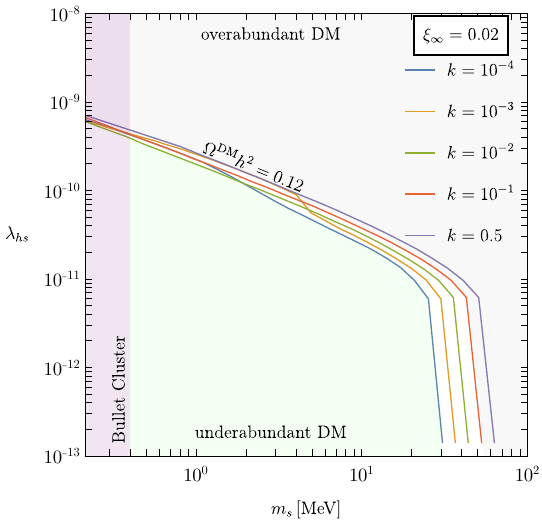}
    \caption{Colored lines display the solutions of the observed relic abundance of DM~\cite{Planck:2018vyg} for different values of $k$ in the parameter plane $\lambda_{hs}\  \text{vs}\ m_s$ with $\lambda_s=10^{-2}$ and $\xi_\infty = 10^{-3}$ ($\xi_\infty=0.02$) in the left (right) plot. The green (gray) region results in an under (over) abundance of DM and depend on the initial condition, $\xi_\infty$. The purple region is excluded by DM self-interactions. Note that the solutions do not form a straight line in the parameter plane; certain regions along the solution line exhibit smaller values of $\lambda_{hs}$. This is a result of $2\to 3$ self-interactions overproducing DM, counterbalanced by lowering the value of the portal coupling.}
    \label{fig:resultsZ3}
\end{figure}
In order to study the interplay of freeze-in and cannibal reactions, and consequently determine the values of $\lambda_{hs}$ that lead to the observed relic abundance, we perform a parameter scan in the plane $\lambda_{hs}\ \text{vs}\ m_s$ for different choices of $k(=g_s^2/(3\lambda_s m_s^2))$, as depicted in Figure~\ref{fig:resultsZ3}. Colored lines display solutions corresponding to the observed DM relic abundance, fixing $\lambda_s = 10^{-2}$. The purple region is excluded by too strong self-scatterings using Eq.~\eqref{eq:sigmaT}, where $\sigma_T$ encompasses both $SS\to SS$ and $SS^*\to SS^*$. Note that this constraint depend on $m_s$, $\lambda_s$ and $k$: for $m_s = 400\,\text{keV}$, $\lambda_s = 10^{-2}$ and $k=10^{-4}$ ($k=0.5$), $\sigma_T/m_s=0.54\,\text{cm}^2/\text{g}$ ($\sigma_T/m_s=0.64\,\text{cm}^2/\text{g}$). The displayed exclusion region corresponds to the most stringent constraint.

Let us first discuss the impact of the strength of number changing self-interactions by comparing lines for different values of the $k$ parameter focusing first on the left panel of Figure~\ref{fig:resultsZ3}. For $m_s\lesssim 1\,\text{MeV}$ the dark sector successfully reaches chemical equilibrium at high temperatures, subsequently undergoing the standard cannibalization phase. If self-interactions increase (by increasing $k$), more DM is depleted during the dark freeze-out. This depletion is compensated by initially producing more DM (increasing $\lambda_{hs}$). The second region of interest is for masses in the interval $1\,\text{MeV}\lesssim m_s\lesssim 1\,\text{GeV}$. Here the dark sector struggles to attain chemical equilibrium. The non-monotonic behaviour of the solution lines in this mass range is due to DM self interactions decoupling \textit{before} reaching chemical equilibrium. As a result, the cannibalization phase is not achieved and no DM is depleted by $3\to 2$ reactions, thereby necessitating a lower FI production. In the third region, $m_s\gtrsim 1\,\text{GeV}$, self-interactions are negligible (cf. Eq.~\eqref{eq:ap_m_scaling}) to have any sizeable impact on $\lambda_{hs}$. 

Now let us turn to the impact of $\xi_\infty$, by comparing the left and right plots in Figure~\ref{fig:resultsZ3}. For $\xi_\infty = 10^{-3}$, i.e. very small initial abundance of DM, a somewhat larger values of $\lambda_{hs}$ are required compared to the $\xi_\infty=0.02$ case. Also the interplay between the number changing self-interactions and freeze-in contribution, seen as departure from simple power-law scaling, is more pronounced for lower values of $\xi_\infty$, since otherwise the kinetic energy transferred to the dark sector during FI is comparable to its already existing energy density. 

To summarize, these results offer an insight on the rich dynamical interplay of the SIDM+FIMP scenarios and its effect on the parameter space. Additionally, it is worth noting that having $\xi_\infty<1$ allows the $\mathbb{Z}_3$ model to achieve the SIDM realisation which is consistent with relic abundance and without conflict with structure formation, which was found to be not possible for $\xi_\infty=1$ \cite{Choi:2016tkj}. However, this particular realization is beyond the reach of current experimental prospects. 
If the dark sector consists of only one species of particles that are singlets with respect to the SM gauge group, any detection becomes challenging because all non-gravitational interactions are mediated only by the Higgs boson. Among the future prospects, there are numerous ongoing experimental efforts to detect sub-GeV dark matter. Outstanding example are the DM-electron scattering DD experiments targeting DM masses in the MeV range~\cite{Essig:2011nj,Derenzo:2016fse,Essig:2017kqs}. Currently, telescope observations also aim at targeting such masses~\cite{Kumar:2018heq}. Interestingly, masses in the GeV range are already testable in DD experiments~\cite{Hambye:2018dpi}. Our interest, however, lies in cannibal sectors, thus masses in the MeV range are our primary focus. Noteworthy, electron recoil experiments are currently unable to test FI couplings in this mass range.\footnote{The non-relativistic DM-electron scattering cross section for this HP model is given roughly by $\sigma_{S e\to Se}\sim \lambda_{hs}^2 m_e^2/m_h^4\sim 4\lambda_{hs}^2\times 10^{-43}\,\text{cm}^2$. The sensitivity of silicon based electron recoil experiments is $\sigma_{\text{DM}\,e\to\text{DM}\,e}\sim10^{-42}\,\text{cm}^2$ for $m_\text{DM}\sim 10\,\text{MeV}$~\cite{Ramanathan:2020fwm}.}

\section{SM $+$ $\mathbb{Z}_3$ complex scalar $+$ scalar mediator}
\label{sec:Z3_w_mediator}

Finally, in this section we discuss our main model incorporating all the interesting dynamics of the simpler models presented before while offering richer phenomenology. This model is defined by extending the potential~\eqref{HP_Z3} by an additional real singlet scalar field $\phi$, which interacts with both the Higgs doublet of the SM and the complex scalar DM. 

\subsection{The model}

The generalized renormalizable potential can be expressed as:
\begin{equation}\label{eq:full_potential}
    V(H,S,\phi)=V_s + V_\phi +V_{\phi s} + V_\text{HP} \,,
\end{equation}
where
\begin{equation}
    \begin{split}
        V_{\phi} &=\lambda_1\phi+\frac{\mu_\phi^2}{2}\phi^2 + \frac{g_\phi}{3!}\phi^3 + \frac{\lambda_\phi}{4!}\phi^4\,,
        \\V_{\phi s} &= \frac{\lambda_{\phi s}}{2}\,\phi^2\vert S\vert^2 + A_{\phi s}\,\phi \vert S\vert^2 + \kappa_{\phi s}\,\phi S^3 + \text{c.c.}\,,
        \\V_\text{HP} &= \left(\lambda_{h s}\,\vert S\vert^2 + \lambda_{h\phi}\,\phi^2 + B_{h\phi}\,\phi\right)\vert H\vert^2\,,
    \end{split}
\end{equation}
and $V_s$ is defined in Eq.~\eqref{eq:VS}, while $S$ and $\phi$ can exchange entropy through $V_{\phi s}$ (see e.g.~\cite{Ghosh:2020lma}). Here $\kappa_{\phi s}$ can lead to the FI semi-production of DM via $\phi S\to S S$~\cite{Bringmann:2021tjr,Hryczuk:2021qtz}. In this study we choose to neglect this interaction through setting $\kappa_{\phi s}=0$ and focus on the thermalization of the DM and the mediator through $\lambda_{\phi s}$. To~streamline the analysis, we also set $\lambda_{hs} = 0=\lambda_{h\phi}$, considering creating the population of the dark sector via $B_{h\phi}$. In the forthcoming discussion, we assume the stability of $S$ ($k<8/3$, where $k=g_s^2/(3\lambda_s \mu_s^2)$). The mediator mixes with the Higgs via $B_{\phi h}$, which is parameterized by the mixing angle $\theta$ given in Eq.~\eqref{eq:mixing_angle} with the detailed analysis of the stability of the potential provided in the Appendix~\ref{ap:vac_stab}.

\subsection{Relevant reactions and cBEs}
%
\begin{table}[t!]
    \centering
    \begin{tabular}{|c|c|l|}
        \hline
        Process & Interaction term & Description \\[0.5ex]
        \hline\hline
        $\phi\phi\leftrightarrow SS^*$  & $\lambda_{\phi s}\,\phi^2\vert S\vert^2$ & production-annihilation \\[1ex]
        $\phi S^{(*)}\leftrightarrow \phi S^{(*)}$ & $\lambda_{\phi s}\,\phi^2\vert S\vert^2$ & elastic scattering \\[1ex]
        $h\to SS^*$ & $A_{\phi s}\theta \,h\vert S\vert^2$ & dominant freeze-in \\[1ex]
        $h\to \phi SS^*$ & $\lambda_{\phi s}\theta \,h\phi\vert S\vert^2$ & sub-dominant freeze-in\\[1ex]
        $2\leftrightarrow 3$ & $g_sS^3,\,\lambda_s\vert S\vert^4$ & cannibalization \\
        $hh\to h\phi$ & $-\lambda_h\theta \phi h^3$ & sub-dominant freeze-in for $\phi$ \\
        \hline
    \end{tabular}
    \caption{List of most relevant processes that impact the DM and mediator abundance. Remaining processes, involving electroweak bosons, are given in the Appendix~\ref{ap:EW}.}
    \label{tab:processes}
\end{table}

The DM and mediator abundances are impacted by the processes shown in Table~\ref{tab:processes}. The mixing between the scalar field and the Higgs boson post-EWPT leads to the substitution $\phi \to \cos\theta\,\phi + \sin\theta\,h\approx \phi + \theta\,h$, inducing interactions between the DM and the Higgs boson via interactions with $\phi$:
\begin{equation}\label{eq:terms_in_Vphis}
\begin{split}
V_{\phi s} \supset\lambda_{\phi s}\theta\,\phi h \vert S\vert^2  + A_{\phi s}\theta\,h\vert S\vert^2 + \mathcal{O}(\theta^2)\,.
\end{split}
\end{equation}
The first term induces three-body Higgs decay, $h\to \phi SS^*$ when kinematically allowed. As $m_\phi,\,m_s\ll m_h$, we estimate this decay in the massless limit for the DM and mediator within the cBE. The Higgs potential also induces interactions between $\phi$ and the Higgs boson after rotation,\footnote{Through the mixing with the Higgs $V_\phi$ will lead to similar interactions through $g_\phi$ and $\lambda_\phi$, which we assume to be negligible.} 
\begin{equation}\label{eq:scalar_higgsinteract}
    \lambda_h\, (H^\dagger H)^2 = \frac{\lambda_h}{4}(h-\theta\phi+v)^4 = -\lambda_h\,\theta \,h^{3}\phi+\mathcal{O}(\theta^2)\,.
\end{equation}
Avoiding mediator decay into DM ($m_\phi<2m_s$), the set of cBEs is:
\begin{align}\label{eq:system_med_DM}
        \frac{Y'_S}{Y_S}&=\frac{1}{x\,\tilde H}\left(\braket{C_{h\to \phi SS^*}} + \braket{C_{h\to SS^*}}+\braket{C_{\phi\phi\leftrightarrow SS^*}} +  \braket{C_{3\leftrightarrow 2}} \right)\,,\nonumber
        \\-\frac{x_S'}{x_S}&=
        \begin{aligned}[t]&\frac{1}{x\,\tilde H}\big(\braket{C_{h\to \phi SS^*}}_2+\braket{C_{h\to SS^*}}_2+\braket{C_{\phi S\leftrightarrow \phi S}}_2 +  \braket{C_{3\leftrightarrow 2}}_2 \big)
        \\&-\frac{Y'_S}{Y_S} + \frac{H}{x\,\tilde H}\frac{\braket {p^4/E^3}}{3T_S} + \frac{2s'}{3s}\,,\end{aligned}
        \\\frac{Y'_\phi}{Y_\phi}&=\frac{1}{x\,\tilde H}\left(\braket{C_{h\to \phi SS^*}}+\braket{C_{\text{SM SM}\to \text{SM}\,\phi}}
        +\braket{C_{\phi\phi\leftrightarrow SS^*}}\right)\,,\nonumber
        \\-\frac{x_\phi'}{x_\phi}&=\frac{1}{x\,\tilde H}\left(\braket{C_{h\to \phi SS^*}}_2+\braket{C_{\text{SM SM}\to \text{SM}\,\phi}}_2+\braket{C_{\phi S\leftrightarrow \phi S}}_2  \right) -\frac{Y'_\phi}{Y_\phi} + \frac{H}{x\,\tilde H}\frac{\braket {p^4/E^3}}{3T_\phi} + \frac{2s'}{3s}\,.\nonumber
\end{align}
Here we define $x_S=m_s/T_S$ and $x_\phi=m_\phi/T_\phi$. The collision operators of $S-\phi$ interactions are detailed in the Appendix~\ref{ap:C_DM_med}, while the triple Higgs decay collision term can be found in the Appendix~\ref{ap:triple_higgs_decay}.

\subsection{Results}\label{sec:Z3_w_mediator_results}
This model offers considerably richer dynamics compared to the previous two. We will first address these features showing four examples that illustrate how DM cannibalization is supported or counterbalanced by the mediator, depending on the mass hierarchy between them and the strength of the number changing self-interactions.   

Secondly, we will examine the mediator's phenomenology, addressing mainly the cosmological bounds over the \textit{mediator's} lifetime. Unlike to the secluded $\mathbb{Z}_3$ case of Section~\ref{sec:Z3}, this model predicts testable signals in telescopes and searches for long-lived particles. If the mediator mass lies within the MeV scale its decay can significantly influence the cosmological evolution of the Universe. This impact could potentially alter the predicted abundances of primordial elements produced during BBN ($T\sim 0.1-1\,\text{MeV}$). Such constraints are highly dependent on the abundance of the mediator just before and during the BBN epoch. 

Late mediator decays can inject energy into the SM plasma, potentially disrupting Cosmic Microwave Background (CMB) observations. The production of electron/positron pairs from $\phi$ can also result in the flux of X-ray photons from the Inverse Compton Scattering (ICS) process. In the results below we will present the parameter space that does not conflict with these observations. Finally, we will discuss the detectability of the mediator in the forthcoming long-lived particle search experiment, MATHUSLA.

\subsubsection{Benchmarks solutions of the cBE system}\label{sec:Z3_w_mediator_bms}

\begin{figure}[t!]
    \centering
     \begin{subfigure}
        \centering
        \includegraphics[width=0.46\textwidth]{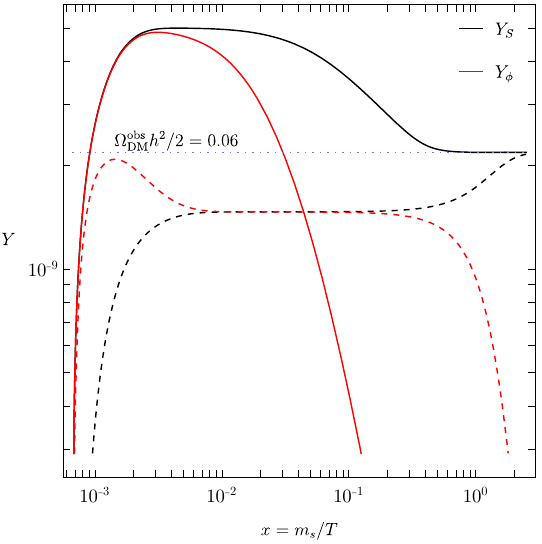}
    \end{subfigure}
    \hfill
    \begin{subfigure}
        \centering
        \includegraphics[width=0.46\textwidth]{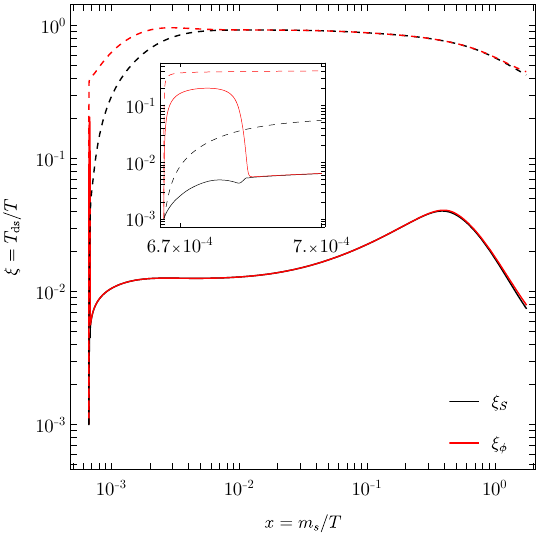}
    \end{subfigure}
    \\
     \begin{subfigure}
        \centering
        \includegraphics[width=0.46\textwidth]{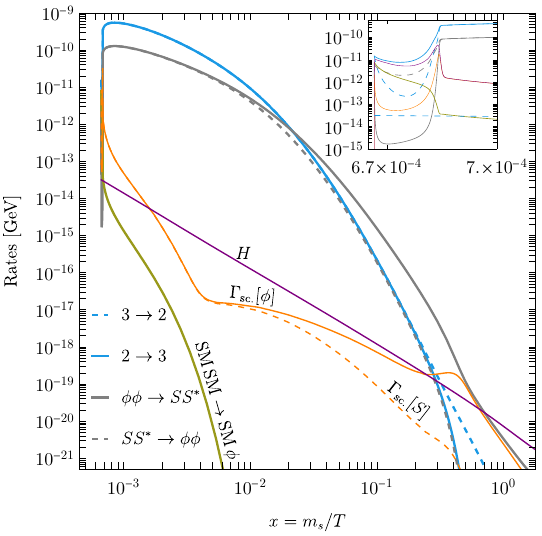}
    \end{subfigure}
    \hfill
    \begin{subfigure}
        \centering
        \includegraphics[width=0.46\textwidth]{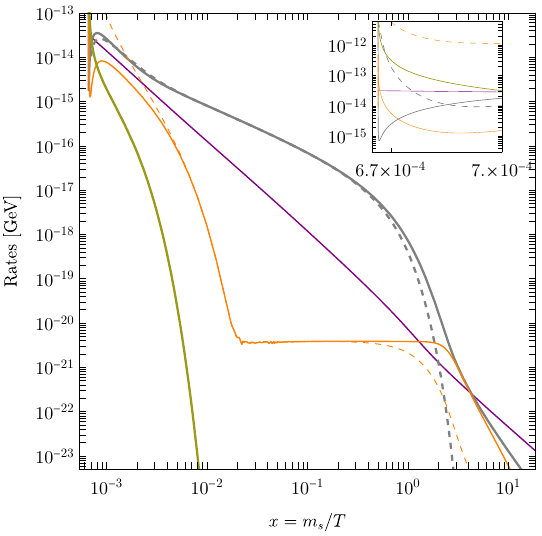}
    \end{subfigure}
    \caption{Evolution of the system for $m_s = 100\,\text{MeV}$ with $m_\phi = 160\,\text{MeV}$, $\lambda_{\phi s} = 10^{-3}$, $A_{\phi s}=0$, $k = 0.5$, $\lambda_s = 0.05 $ and $\theta$ was chosen to match the observed relic abundance (dotted line). Bottom: the right (left) plot display the rates in the case without (with) self-interactions. Top: the black and red dashed (thick) lines display the evolution without (with) self-interactions with $\theta\simeq  2.2\times 10^{-10}$ ($\theta\simeq 5.2\times 10^{-11}$). Inset plots are zooming on the low $x$ region to resolve the effect of sudden injection of particles after EWPT. 
    }
    \label{fig:evol1}
\end{figure}

\begin{figure}[t!]
    \centering
     \begin{subfigure}
        \centering
        \includegraphics[width=0.46\textwidth]{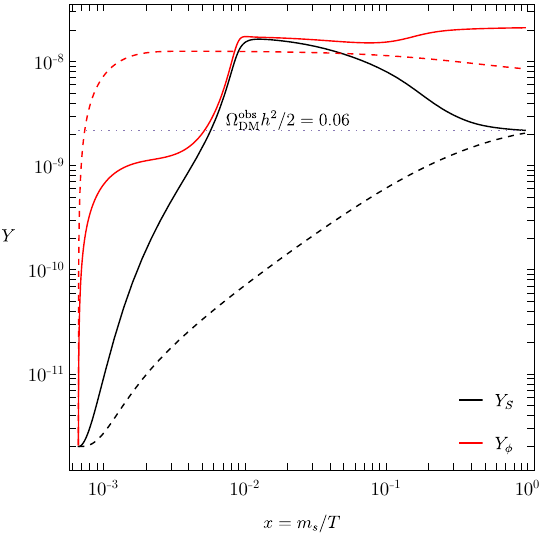}
    \end{subfigure}
    \hfill
    \begin{subfigure}
        \centering
        \includegraphics[width=0.46\textwidth]{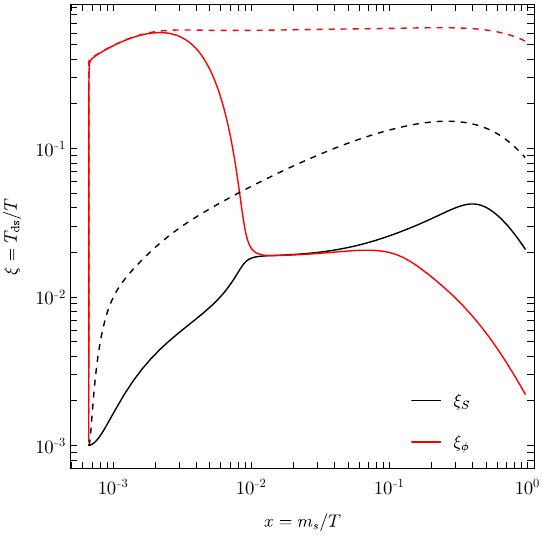}
    \end{subfigure}
    \\
    \begin{subfigure}
        \centering
        \includegraphics[width=0.46\textwidth]{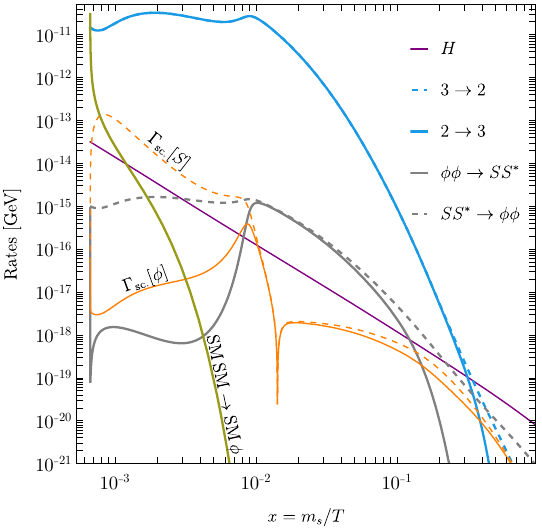}
    \end{subfigure}
    \hfill
    \begin{subfigure}
        \centering
        \includegraphics[width=0.46\textwidth]{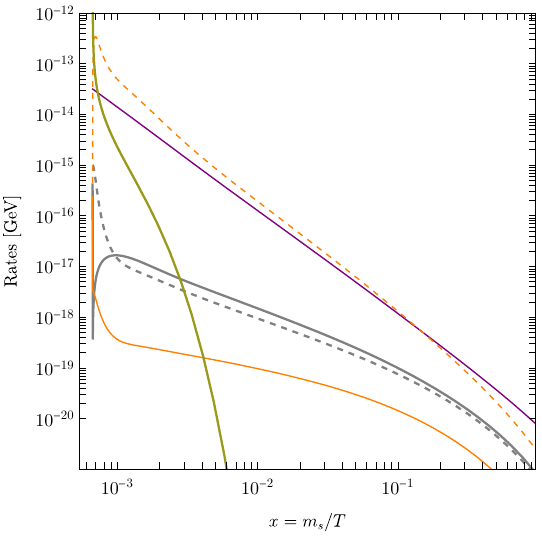}
    \end{subfigure}  
    \caption{Same as Figure~\ref{fig:evol1} with $m_\phi=80\,\text{MeV}<m_s=100\,\text{MeV}$ and $\lambda_{\phi s} = 10^{-5}$. In this scenario $\theta \simeq 3.73\times10^{-10}$ ($\simeq 1.12\times 10^{-10}$) without (with) self-interactions.}
    \label{fig:evol2}
\end{figure}

\begin{figure}[t!]
    \centering
     \begin{subfigure}
        \centering
        \includegraphics[width=0.46\textwidth]{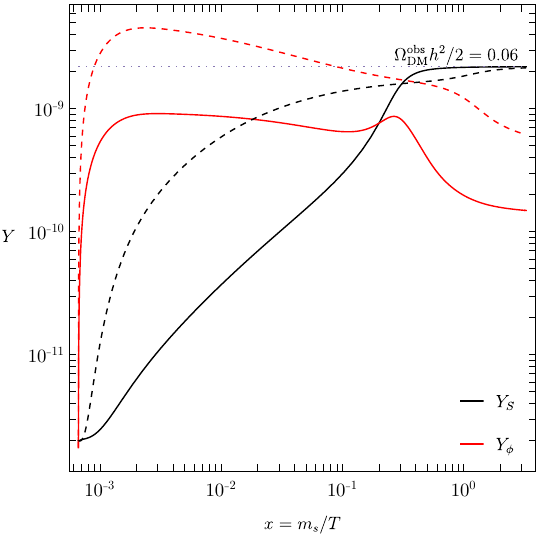}
    \end{subfigure}
    \hfill
    \begin{subfigure}
        \centering
        \includegraphics[width=0.46\textwidth]{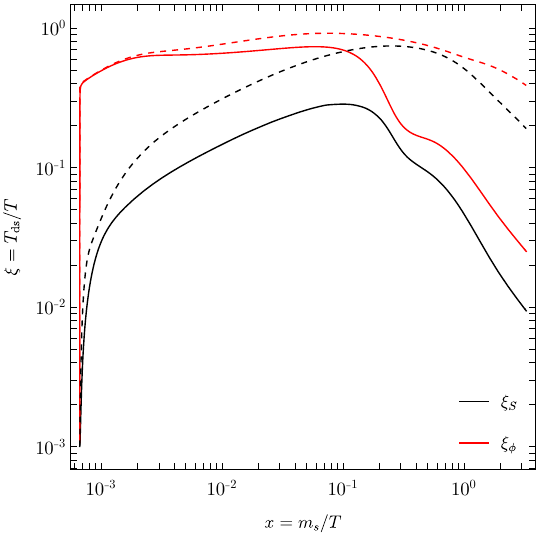}
    \end{subfigure}
    \\
    \begin{subfigure}
        \centering
        \includegraphics[width=0.46\textwidth]{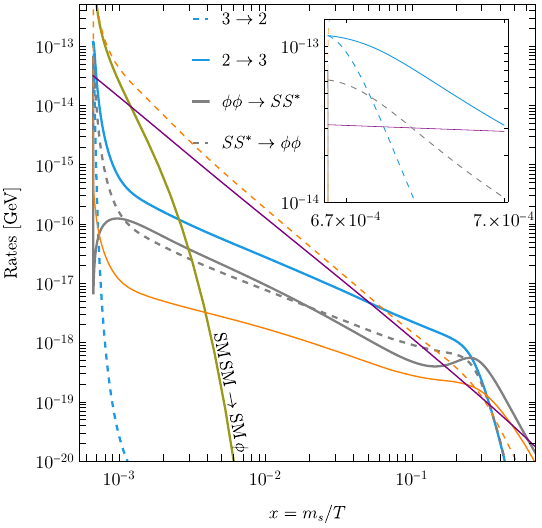}
    \end{subfigure}
    \hfill
      \begin{subfigure}
        \centering
        \includegraphics[width=0.46\textwidth]{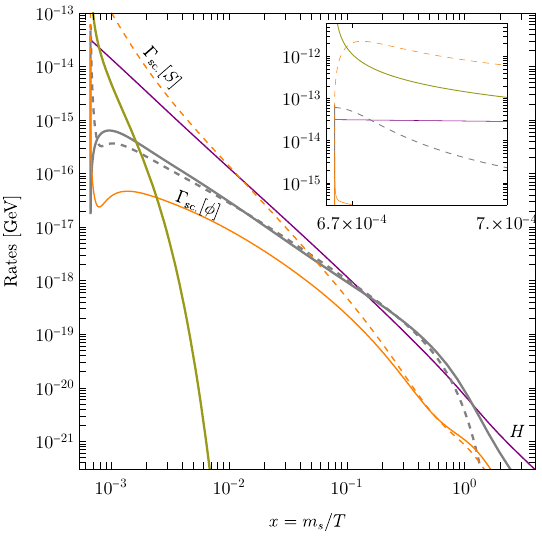}
    \end{subfigure}
    \caption{Same as Figure~\ref{fig:evol1} with $\lambda_{\phi s} = 10^{-4}$ and $\lambda_s = 10^{-2}$. In this scenario $\theta \simeq 2.3\times10^{-10}$ ($\simeq 1\times 10^{-10}$) without (with) self-interactions.}
    \label{fig:evol3}
\end{figure}

\begin{figure}[t!]
    \centering
     \begin{subfigure}
        \centering
        \includegraphics[width=0.46\textwidth]{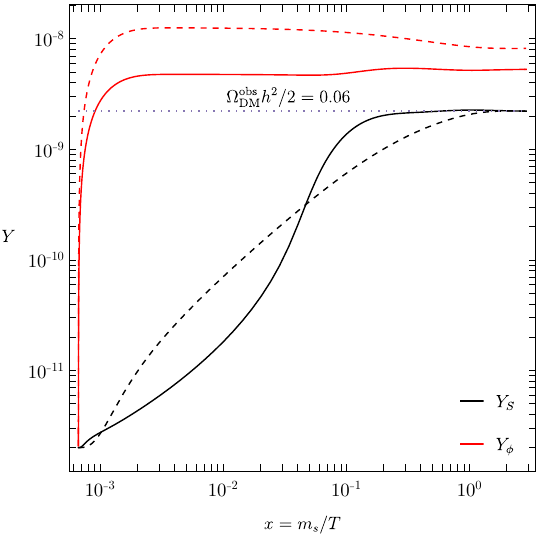}
    \end{subfigure}
    \hfill
    \begin{subfigure}
        \centering
        \includegraphics[width=0.46\textwidth]{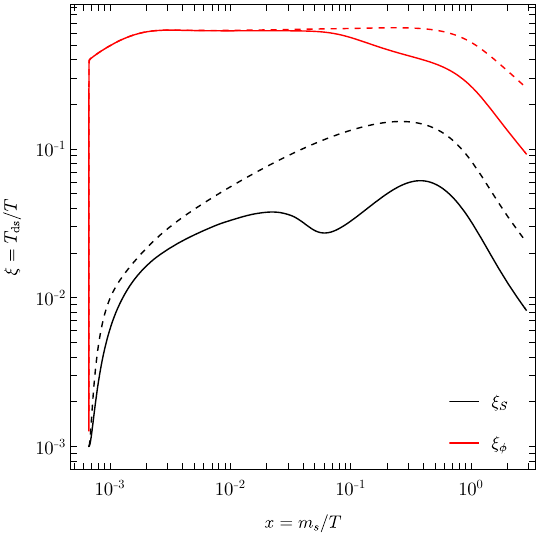}
    \end{subfigure}
    \\
    \begin{subfigure}
        \centering
        \includegraphics[width=0.46\textwidth]{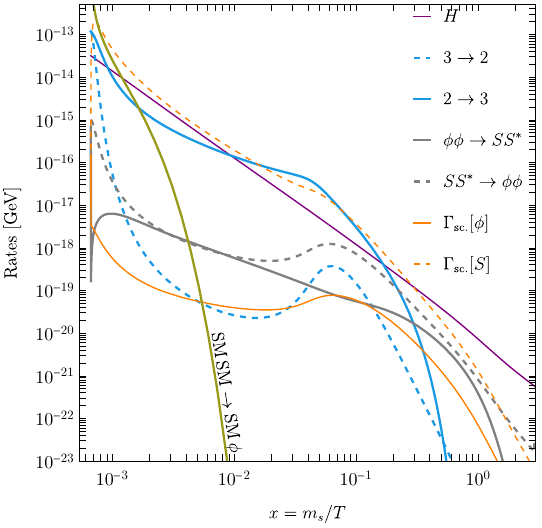}
    \end{subfigure}
    \caption{Same as Figure~\ref{fig:evol2} with $\lambda_s = 10^{-2}$. In this scenario $\theta \simeq 3.73\times10^{-10}$ ($\simeq 2.29\times 10^{-10}$) without (with) self-interactions. The rates without self-interactions are the same as in Figure~\ref{fig:evol2} (bottom right).}
    \label{fig:evol4}
\end{figure}

We start with four benchmark examples  leading to correct DM relic abundance and showcasing distinct dynamics, Figures~\ref{fig:evol1}-\ref{fig:evol4}. The first two cases illustrate the evolution with strong self-interactions where the system quickly achieves equilibrium. The second two cases show the evolution with relatively weak self-interactions. Each figure shows the yield (top left), ratio of temperatures ($\xi=T_\text{ds}/T$, top right), the rates with DM self-interactions (bottom left) and without DM self-interactions (bottom right). The rates without self-interactions, $\lambda_s=0$, correspond to the dashed lines in the evolution equations. The rates are evaluated at the zeroth moment level, $\Gamma_\text{rate} = \braket{C_\text{int.}}$, with the exception of the thick and dashed orange lines, which represent the DM-mediator scattering rate, 
\begin{equation}
    \begin{split}
        \Gamma_\text{sc.}[\phi] &= \vert\braket{C_\text{scatter}[\phi]}_2\vert\,,
        \\\Gamma_\text{sc.}[S] &= \vert\braket{C_\text{scatter}[S]}_2\vert\,,
    \end{split}
\end{equation}
where $\braket{C_\text{scatter}}_2$ is given in Eq.~\eqref{eq:S_med_scatter}.

Let us now discuss each case in detail. First, for $m_s<m_\phi$ with strong interactions. The evolution is depicted in Figure~\ref{fig:evol1} (top). The first stage of the evolution is characterized by an instantaneous injection of energy at $T=150\,\text{GeV}\,(x\simeq 6.6\times 10^{-4})$, the rate of FI production is shown in dark green lines in the rate plots (bottom). Note that the mediators are primarily populated via $\text{SM}\,\text{SM}\to \text{SM}\,\phi$ and they quickly dissipate part of this energy into the DM fluid through scattering in a sudden attempt to achieve kinetic equilibrium (see inset plot). In this case, the DM-mediator interactions are sufficiently strong ($\lambda_{\phi s}=10^{-3}$) to bring both fluids into kinetic and chemical equilibrium in the early stages of the evolution. Meanwhile, DM self-interactions re-establish chemical equilibrium in the DM fluid ($\lambda_s = 0.05$, cf. thick and dashed lines in the rate plots). As the system evolves, mediators rapidly deplete to produce more DM (thick gray line in bottom left plot), counteracting the onset of the cannibalization phase.

Secondly, $m_s>m_\phi$ with weak interactions within the dark sector. Here there are two sources of DM depletion, namely self- ($3\to 2$) and DM-mediator ($SS^*\to \phi\phi$) annihilations, and the evolution is depicted in Figure~\ref{fig:evol2} (top). A similar study has been conducted in~\cite{Dey:2016qgf}. To ensure the observed DM relic abundance, the interactions in the dark sector must be weaker compared to the previous scenario (hence $  \lambda_{\phi s} = 10^{-5}$ here). At the same time, DM self-interactions drive the DM fluid back into chemical equilibrium, which influences the evolution of the mediators, as they are partially coupled to the DM during the evolution (thick and dashed gray/orange lines around $x= 10^{-2}$ in the bottom left plot). While both fluids try to reach equilibrium, more mediators are produced from DM, which are colder than those produced by freeze-in, thus cooling the mediator sector. In this instance, $\Gamma_\text{sc.}[S]$ (orange dashed) is the rate of heat injected from the mediator fluid, while $\Gamma_\text{sc.}[\phi]$ (orange solid) is the rate of heat loss before an equilibrium is reached.
    
Third, an unsuccessful attempt of the dark sector's interactions to achieve equilibrium for $m_s<m_\phi$. This is depicted in Figure~\ref{fig:evol3}, and unlike in the last two cases, there are no abrupt attempts to re-establish equilibrium. As the universe expands, the $2\to 3$ reaction becomes more efficient around $x=3\times 10^{-2}$ (blue thick line), driving the DM fluid toward chemical equilibrium by producing more dark mater at the cost of kinetic energy. This DM population possesses enough energy to produce more mediators (dashed gray line bottom left). However, since mediators are heavier, this process quickly becomes inefficient, and the reverse process rapidly dominates (thick vs dashed gray lines bottom left). The system decouples before reaching equilibrium. 
    
Finally, the last case involves weak self-interactions with $m_s>m_\phi$ and is shown in Figure~\ref{fig:evol4}. The primarily active reactions are $2\to 3$ and $\phi S\leftrightarrow\phi S$. The entire production of DM is attributed to the $2\to 3$ reaction, which initially surpasses the rate of expansion but quickly drops below the Hubble rate, remaining so until $x\sim 2\times 10^{-2}$ when its efficiency increases again (cf. blue thick and orange dashed lines). In fact, the rate $\Gamma_{2\to 3}$ remains fairly efficient due to the mediator sector supplying heat to the DM fluid through scattering, which is subsequently transformed into number density by the $2\to 3$ reaction. Concurrently, number changing self-interactions keep DM cool (cf. thick and dashed black lines in the temperature plot), particularly when $\Gamma_{2\to 3}>\Gamma_\text{sc}[S]$ around $x=3\times 10^{-2}$ (bottom), thus decreasing the DM temperature, while its yield is further boosted, followed by an increase of the scattering rate that injects more heat into DM around $x=10^{-1}$. At this stage, $\Gamma_{SS^*\to \phi\phi}$ starts growing as the DM sector possesses enough energy to produce mediators (cf. gray dashed), thereby increasing the mediator's abundance.

\subsubsection{Numerical scan setup}\label{sec:Z3_w_mediator_scan}
%
To investigate the phenomenological implications of this model, we performed a random scan across the parameters outlined in Table~\ref{tab:params_scan}, fitting to the value of the observed relic abundance, i.e., accepting points within $\Omega_\text{DM}^\text{obs}/\Omega_\text{DM}=1\pm 0.01$, in accordance with the parameter 68\% intervals from the  TT,TE,EE+lowE+lensing data as reported by the Planck Collaboration~\cite{Planck:2018vyg}.
In order to study the implications of various strengths of DM self-interactions, we fixed $k=0.5$,\footnote{Note that both $k$ and $\lambda_s$ influence the strength of self-interactions, and one can compensate for the other. Also note that $k=0.5$ is safely within the stability bound ($k<8/3$).} and allowed $\log_{10}\lambda_s$ to take values on a grid from -4 to -1 with a step of 0.5.
\begin{table}[t!]
\begin{center}
{
   \begin{tabular}{|c|c|c|c|}
   \cline{1-4}
    Parameter  & min. & max. & Impact on \\
     \hline\hline
   \raisebox{-0.5ex}{$m_s$} & \raisebox{-0.5ex}{$500\,\text{keV}$} & \raisebox{-0.5ex}{$10\,\text{GeV}$} & \raisebox{-0.5ex}{ID and DD}  \\[5pt]
   \cline{1-4}
   \raisebox{-0.5ex}{$r=\frac{m_\phi}{2m_s}$} & \raisebox{-0.5ex}{0} & \raisebox{-0.5ex}{1}  & \raisebox{-0.5ex}{ID and DD} \\ [5pt]
   \cline{1-4}
  \raisebox{-0.5ex}{ $\tilde A=A_{\phi s}/m_\phi$}  & \raisebox{-0.5ex}{$10^{-7}$} & \raisebox{-0.5ex}{$1$}  &  \raisebox{-0.5ex}{ID and FI }    \\ [5pt]
   \cline{1-4}
   \raisebox{-0.5ex}{$\lambda_{\phi s}$} & \raisebox{-0.5ex}{$10^{-10}$} &  \raisebox{-0.5ex}{$10^{-1}$}  &  \parbox[t]{8cm}{ID, DD, FI and equilibrium between $\phi$ and $S$}   \\ 
   \cline{1-4}
   \raisebox{-0.5ex}{$\log_{10}\lambda_{s}$} & \raisebox{-0.5ex}{$-4$} &  \raisebox{-0.5ex}{$-1$}  &  \raisebox{-0.5ex}{DM self-interactions}   \\ 
   \cline{1-4}
    \end{tabular}
    \caption{Relevant parameters, their ranges in the numerical scan and phenomenological impact. The choice of the DM mass scale is motivated by $3\leftrightarrow 2$ processes and that for $m_s\lesssim \mathcal{O}(100\,\text{keV})$ the bounds on self-scatterings (Eq.~\eqref{eq:sigmaT}) are very stringent. The choice of values for $\tilde A$ and $\lambda_s$ is to preserve unitarity and perturbativity~\cite{Schuessler:2007av}. The listed parameters are scanned with $\log_{10}$ scaling, with the exception of $r$, which is scanned with a linear scaling.}
    \label{tab:params_scan}
}
\end{center}
\end{table}

The phenomenology of the model branches into two distinct cases based on the mass hierarchy between $S$ and $\phi$. For one, only the $m_s>m_\phi$ case leads to any appreciable DM annihilation signals, as the $\phi$ mixing with the Higgs is too weak for any direct process of $SS \to \text{SM}$ to be strong enough: only annihilation $SS \to \phi\phi$ followed by the decay of $\phi$'s can have appreciable cross section. But also the mass hierarchy strongly affects the evolution during DM production impacting not only the abundance of $S$, setting the DM relic density, but also the one of $\phi$ altering the total energy injection from its decay, which is an important factor in the determination of the limits from BBN.

Indeed, the presence of $\phi$ could spoil BBN observations if its lifetime is within the range of $10^{-2}\,\text{s}\lesssim \tau_\phi\lesssim10^{12}\,\text{s}$.  Decays at early times ($\tau_\phi\lesssim 10^5\,\text{s}$) are constrained by the refined measurements of the abundance of $^4\,\text{He}$, $^2\,\text{H}$, $^{6}\text{Li}$ and $^7\text{Li}$ as mesons produced by $\phi$ would strongly interact with nucleons. While electromagnetic decays may impact BBN only at comparatively later times ($\tau_\phi \gtrsim10^5\,\text{s}$), mainly due to energetic photons and $e^{\pm}$ interacting efficiently with the background photons, rapidly producing $\gamma$-rays. If such $\gamma$-rays have energies below a certain threshold, they are less likely to interact further with the plasma. This threshold energy changes over time as the universe cools down and is $E\approx 2.2\,\text{MeV}$ at $\tau_\phi\sim10^5\,\text{s}$ and $E\approx 19.8\,\text{MeV}$ at $\tau_\phi\sim10^7\,\text{s}$. These specific values for threshold energies are critical because they mark the points at which $\gamma$-rays photodisintegrate $^2\text{H}$ and $^{4}\text{He}$~\cite{Jedamzik:2006xz}. Therefore, when such additional electromagnetic particles are introduced into the plasma, they immediately trigger an electromagnetic cascade through interactions with the thermal background species. The photons produced in this process subsequently participate in photodisintegration reactions (e.g., $^2\text{H}\,\gamma \to np,$ $^{3}\text{He}\,\gamma \to npp$, etc), leading to a late-time alteration of light-element abundances following nucleosynthesis. To accurately quantify this effect, solving the Boltzmann equations for nuclear abundances is required~\cite{Hufnagel:2020nxa}. However, as this lies beyond the scope of this work, we estimate the BBN constraints using~\cite[Figure 6]{Depta:2020zbh}.

\subsubsection{Scan results for $m_s>m_\phi$}\label{subsubsec:mphi>ms}
\begin{figure}[t!]
    \centering
    \begin{subfigure}
        \centering
        \includegraphics[width=0.48\textwidth]{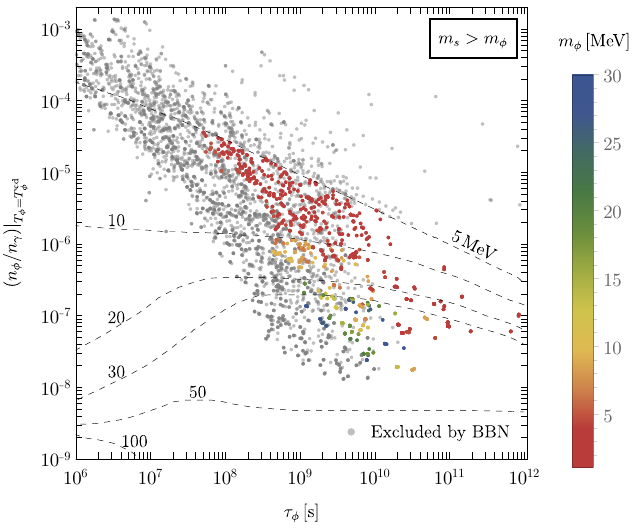}
    \end{subfigure}
    \hfill
    \begin{subfigure}
        \centering
        \includegraphics[width=0.48\textwidth]{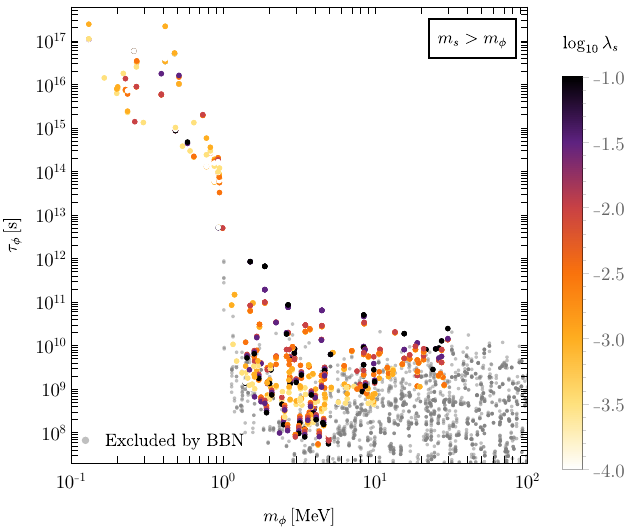}
    \end{subfigure}
    \caption{Points reproducing the observed relic projected onto the planes $\left.n_\phi/n_\gamma\right\vert_{T_\phi=T_\phi^\text{cd}}\,\text{vs}\,\tau_\phi$ (left) and $\tau_\phi\,\text{vs}\,m_\phi$ (right) for masses up to $2m_\mu$, fixing $\xi_\infty = 10^{-3}$ and $k=0.5$. The dashed lines in the left plot indicate the BBN limits for each scalar mass and are adapted from~\cite[Figure 6]{Depta:2020zbh}, while the color bar indicates the mediator mass. There is an overabundance of mediators arising from $S^*S\to 2\phi$, thus mediator masses in the interval $30\,\text{MeV}\lesssim m_\phi<2m_\mu$ are excluded. The allowed points can lead to signals in indirect detection via scalar decay (cf. Figure~\ref{fig:cross_ID}). The right plot includes points with $m_\phi < 1\,\text{MeV}$, which are not in conflict with BBN because their lifetimes exceed $\tau_\phi > 10^{12}\,\text{s}$ with the color bar displaying the values of DM self-interacting coupling $\log_{10}\lambda_s$.
    }
    \label{fig:tauphi_vs_abundance_ms>mphi}
\end{figure}
 \begin{figure}[t!]
 \centering
   \begin{subfigure}
        \centering
        \includegraphics[width=0.49\textwidth]{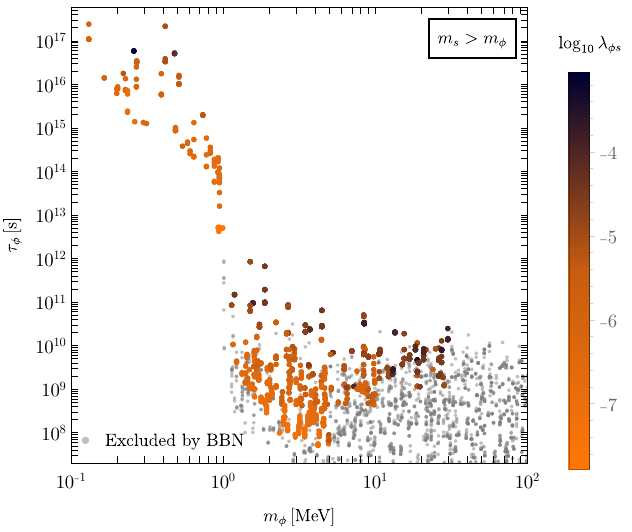}
    \end{subfigure}
    \hfill
    \begin{subfigure}
        \centering
        \includegraphics[width=0.49\textwidth]{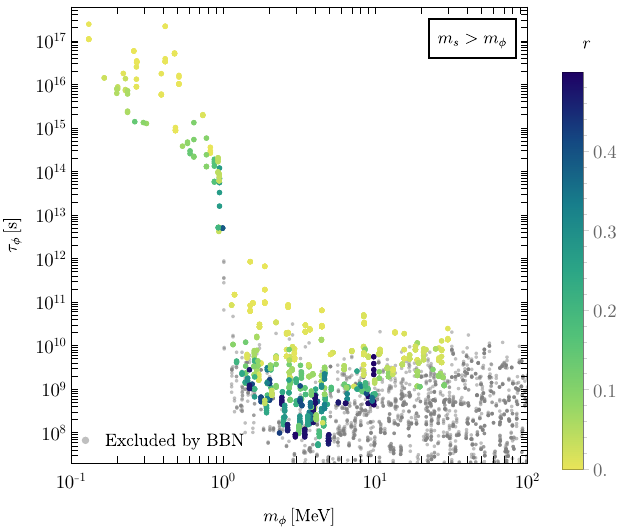}
    \end{subfigure}
    \caption{The same points as in Figure~\ref{fig:tauphi_vs_abundance_ms>mphi} projected on the $\tau_\phi\,\text{vs}\,\,m_\phi$ plane with the color bar displaying the values for $\log_{10}\lambda_{\phi s}$ (left) and $r=m_\phi/(2m_s)$ (right).}
    \label{fig:tauphi_vs_abundance_ms>mphi2}
\end{figure}

\begin{figure}[t!]
    \centering
     \begin{subfigure}
        \centering
        \includegraphics[width=0.48\textwidth]{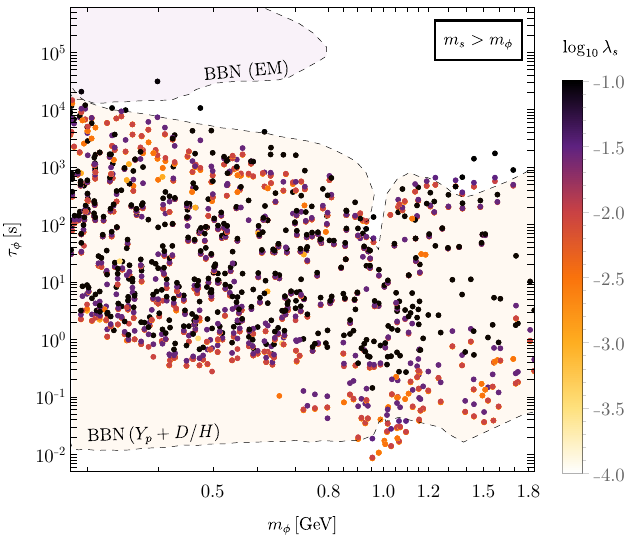}
    \end{subfigure}
    \hfill
    \begin{subfigure}
        \centering
        \includegraphics[width=0.49\textwidth]{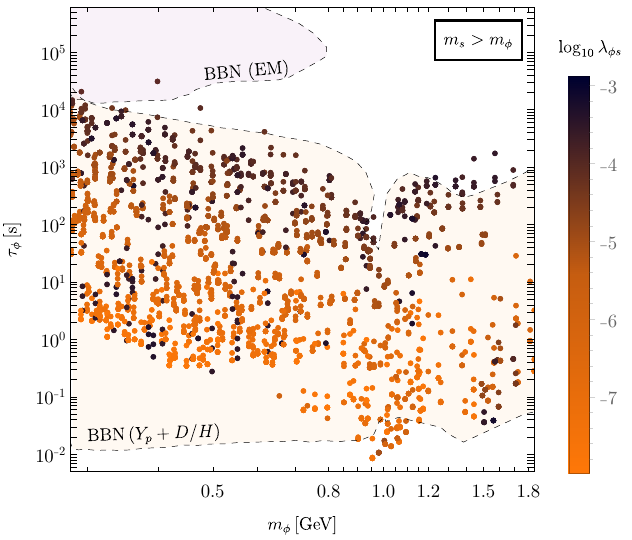}
    \end{subfigure}
     \caption{Same as Figure~\ref{fig:tauphi_vs_abundance_ms>mphi} for mediator masses in the GeV range with the superimposed BBN limits adapted from~\cite[Figure 13]{Fradette:2018hhl}. Note that most of the points are excluded by BBN observations of the primordial mass fraction $Y_p$ and the $D/H$ ratio.}
    \label{fig:tauphi_vs_abundance_ms>mphiGeV}
    \end{figure}

In the case of $S$ heavier than $\phi$ the BBN bounds exacerbate due to the additional production of $\phi$ from $S$, leading to a temporary overabundance of mediators. This case is presented in~\Cref{fig:tauphi_vs_abundance_ms>mphi,fig:tauphi_vs_abundance_ms>mphi2}. All the shown points satisfy the relic density constraint. The ones that are shaded in gray are excluded by the requirement of successful BBN, following the exclusion lines of Figure~\ref{fig:tauphi_vs_abundance_ms>mphi} (left). The remaining points are colored according to the value of a given parameter, chosen differently for each plot to highlight its relevance.

The right plot in Figure~\ref{fig:tauphi_vs_abundance_ms>mphi} and the plots in Figure~\ref{fig:tauphi_vs_abundance_ms>mphi2} show the viable points in the plane of the mediator parameters $\tau_\phi$ vs $m_\phi$. These three plots highlight patterns in different model parameters: $\lambda_s$, $\lambda_{\phi s}$ and $r$, respectively. The lifetime increases with increasing $\lambda_{\phi s}$, while there is no clear pattern when varying $\lambda_{s}$. The mass hierarchy and the DM-mediator coupling (see  Figure~\ref{fig:tauphi_vs_abundance_ms>mphi2}), exhibit a relation that can be understood as follows: if the masses within the dark sector are sufficiently hierarchical ($r\ll 1/2$), then the rate of annihilation $SS^*\to2\phi$ becomes efficient enough that to avoid complete depletion of DM one needs to compensate by lowering $\lambda_{\phi s}$.

Next we turn to mediator with mass exceeding the $\mu^+\mu^-$ threshold. The resulting points satisfying the relic density constraint with superimposed exclusion limits from BBN are shown in  Figure~\ref{fig:tauphi_vs_abundance_ms>mphiGeV}. When $m_\phi>2m_\mu$ rapid mediator decays may successfully evade CMB bounds, but they could still modify the standard BBN predictions of the primordial mass fraction of $Y_p$ and the ratios of primordial number densities $D/H$. The region of the parameter space that evade this constraints corresponds to $m_\phi\gtrsim 1\,\text{GeV}$ with preferred self-interaction coupling $\log_{10}\lambda_{s}\lesssim -2.5$ (top left), while DM-mediator interactions in the range of values $\log_{10}\lambda_{\phi s}\lesssim -5$ and $\log_{10}\lambda_{\phi s}\sim-3$ (top right). 

\begin{figure}[t!]
    \centering
   \includegraphics[width=0.65\textwidth]{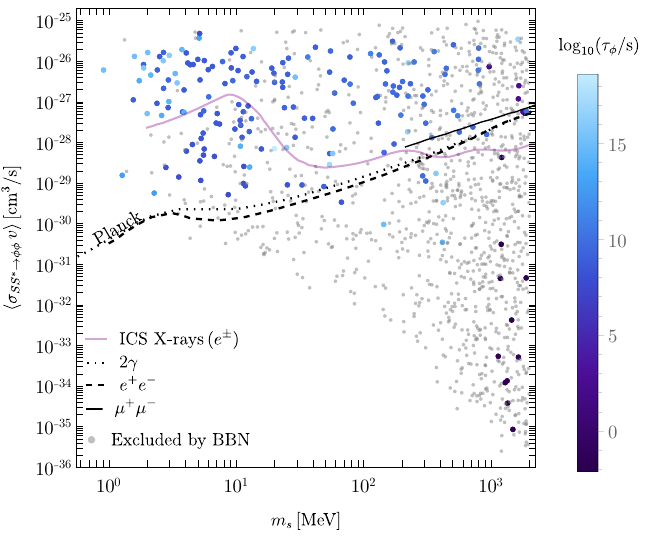}
    \caption{Present day cross section for $SS^*\to 2\phi$ with the superimposed limits of PLANCK~\cite{Planck:2018vyg} (black lines) and from Inverse Compton Scattering (ICS) X-rays produced by energetic electron/positron pairs~\cite{Cirelli:2024kph} (purple line).}
    \label{fig:cross_ID}
\end{figure}

Finally, this scenario is also constrained by the energy injection to the CMB from DM annihilation and X-ray telescopes data. Comparing to the limits set by the PLANCK satellite~\cite{Planck:2018vyg}, recast using~\cite{Slatyer:2015jla}, as well as Inverse Compton Scattering (ICS) X-ray limits, recast using~\cite{Cirelli:2024kph}, in Figure~\ref{fig:cross_ID} we project the otherwise viable points onto the plane of present day $SS^*\to 2\phi$ cross section vs. the DM mass, with the color bar indicating the mediator's lifetime. The limits are given for an idealized case of 100\% branching ratio of $\phi$ to photons, electrons or muons, which are however close enough to make the difference between them to have virtually no impact on the final conclusions. 

To relate the CMB/X-ray constraints on the mediator's lifetime to the limits for DM mass and cross section we assume that \(SS^*\to 2\phi\) happens at rest (i.e. that \(\phi\) possesses energy of \(E_\phi = m_s\)) and we account for the double flux of SM states, as well as for the rescaling of the DM mass by a factor of \(1/2\). While DM in the MeV scale falls within the telescope's sensitivity, the cross section decreases with increasing DM mass, resulting in points that fall below the sensitivity threshold for heavier $S$, particularly for GeV dark matter.

In summary, after careful implementation of the processes affecting the dynamics of freeze-out we see that for $m_s>m_\phi$ and for points predicting the observed value of relic abundance most of the parameter space is strongly constrained by cosmological and observational data, leaving only small windows in the GeV spectrum to be tested by future generation of telescopes.

\subsubsection{Scan results for $m_s<m_\phi$}\label{subsubsec:mphi>ms}

\begin{figure}[t!]
    \centering
    \begin{subfigure}
        \centering
        \includegraphics[width=0.52\textwidth]{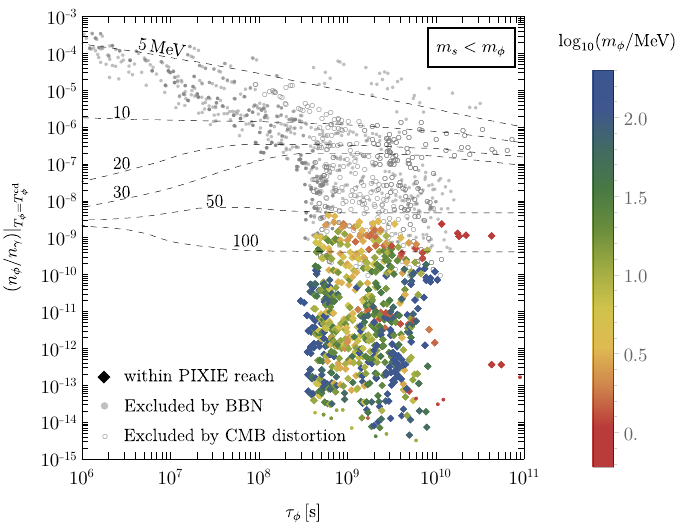}
    \end{subfigure}
    \hfill
    \begin{subfigure}
        \centering
        \includegraphics[width=0.45\textwidth]{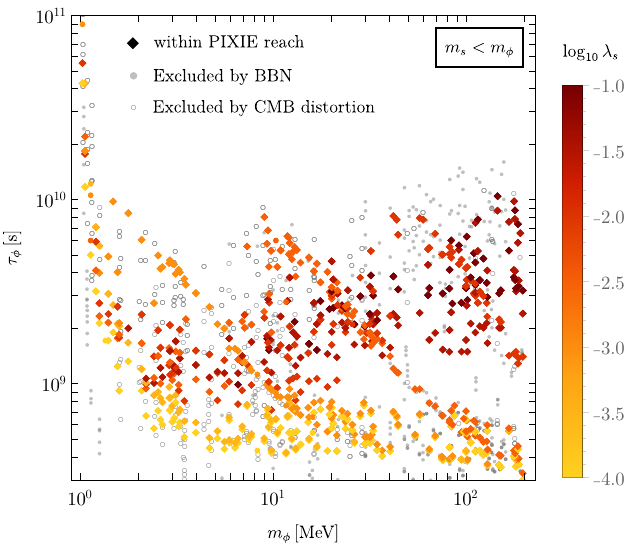}
    \end{subfigure}
    \caption{Same as Figure~\ref{fig:tauphi_vs_abundance_ms>mphi} for the case $m_s<m_\phi$. The empty circles are excluded by CMB distortion. The rhomboids show the points that are within the reach of the PIXIE-like detector~\cite{Kogut:2024vbi}.}
    \label{fig:tauphi_vs_abundance_ms<mphi}
\end{figure}

In contrast, if the hierarchy is inverted, mediators are depleted to produce more DM relaxing the cosmological bounds. This is shown in the left panel of Figure~\ref{fig:tauphi_vs_abundance_ms<mphi}, where a large part of the mediator's abundance lies below the BBN exclusion lines.\footnote{Regarding the BBN bounds for $m_\phi$ above the muon threshold (cf. Figure~\ref{fig:tauphi_vs_abundance_ms<mphi_GeV}), the limits coming from primordial mass fraction and number density $D/H$ require specific calculations for this scenario that go beyond the scope of this work.
}

\begin{figure}[t!]
    \centering
     \begin{subfigure}
        \centering
        \includegraphics[width=0.49\textwidth]{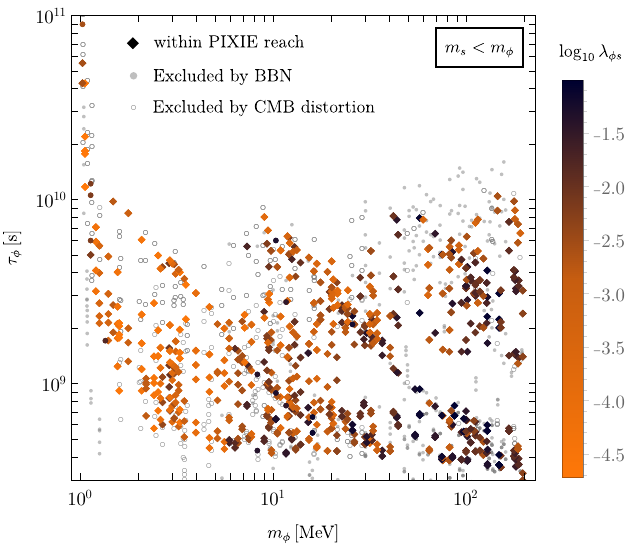}
    \end{subfigure}
    \begin{subfigure}
        \centering
        \includegraphics[width=0.49\textwidth]{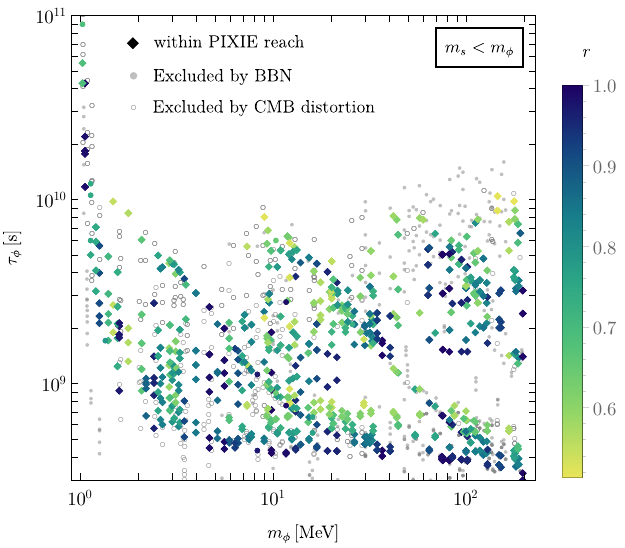}
    \end{subfigure}
    \caption{Same as Figure~\ref{fig:tauphi_vs_abundance_ms>mphi2} with $m_s<m_\phi$.}
    \label{fig:tauphi_vs_abundance_ms<mphi2}
\end{figure}

Unlike in the previous case, the impact of $\phi$ on the CMB spectrum in this scenario requires a more detailed computation of the CMB distortions due to lower abundance of mediators. The photon thermal shape before recombination is measured with high accuracy by COBE/FIRAS~\cite{Fixsen:1996nj}. Decays of $\phi$ around the recombination epoch would inject energy into the photon plasma deviating it from equilibrium and from black body radiation, e.g., late decays can cause changes in the ionaization history around last scattering ($z\simeq 1000$), which in turn would result in changes of the CMB temperature and anisotropies. Distortions can be expressed in the following form~\cite{Chluba:2016bvg}
\begin{equation}
\begin{split}
    y&=\frac{1}{4}\int_{z_\text{rec}}^{z_{\mu y}}\frac{d(Q/\rho_\gamma) }{dz'}dz'\,,
    \\\mu&=1.401\int_{z_{\mu y}}^{\infty}e^{-\left( \frac{z'}{z_\mu}\right)^{5/2}}\frac{d(Q/\rho_\gamma) }{dz'}dz'\,,
\end{split}
\end{equation}
where $z_\text{rec} = 1000$, $z_{\mu y}\simeq 5\times 10^{4}$ and $z_\mu=2\times 10^{6}$.  Points excluded by COBE/FIRAS satisfy~\cite{Fixsen:1996nj}
\begin{equation}
    \begin{split}
        \vert \mu\vert_\text{CF}&<9\times 10^{-5}\,,\\ \vert y\vert_\text{CF}&<1.5\times 10^{-5}\,,
    \end{split}
\end{equation}
and are marked with empty circles in~\Cref{fig:tauphi_vs_abundance_ms<mphi,fig:tauphi_vs_abundance_ms<mphi2}. In future, The Primordial Inflation Explorer (PIXIE) detector~\cite{Kogut:2024vbi} is planned to achieve sensitivity over 1000 times greater than COBE/FIRAS and would test this model if the distortion induced by $\phi$ lies within the ranges:
\begin{equation}
    \begin{split}
        \vert \mu\vert_\text{Pixie}&<10^{-9},
        \\\vert y\vert_\text{Pixie}&<2\times 10^{-9}\,.
    \end{split}
\end{equation}
Points that predict a signal of such strength are marked as rhomboids in the~\Cref{fig:tauphi_vs_abundance_ms<mphi,fig:tauphi_vs_abundance_ms<mphi2,fig:tauphi_vs_abundance_ms<mphi_GeV}. Note that the PIXIE experiment can test most of the points in the MeV range below the muon threshold, with the exception of the ones that fulfill $\left.n_\phi/n_\gamma\right\vert_{\text{cd}}\lesssim 1.63\times 10^{-13}$ (colored filled circles in Figure~\ref{fig:tauphi_vs_abundance_ms<mphi}). Conversely, if the mediator is allowed to decay into muons, its lifetime is shortened, and it falls below the PIXIE sensitivity. Notwithstanding, masses in the interval $[212,284]\,\text{MeV}$ lie within the sensitivity signals from the $\mu$-type distortion (cf. Figure~\ref{fig:tauphi_vs_abundance_ms<mphi_GeV}).

Unlike the previous case, there is an intricate relation between $m_\phi$ and $\tau_\phi$ induced by DM self-interactions that can be appreciated in Figure~\ref{fig:tauphi_vs_abundance_ms<mphi} (right): substantial $\lambda_s$ results in longer lifetimes, as the DM fluid transforms its kinetic energy into number density, thereby necessitating less FI production. Moreover, points that pass the BBN and CMB tests exhibit stronger DM-mediator interactions than in the $m_s>m_\phi$ mass hierarchy. This one can read from Figure~\ref{fig:tauphi_vs_abundance_ms<mphi2} (left), where the allowed region satisfies $-4.5\lesssim\log_{10}\lambda_{\phi s}\le -1$ (in the previous mass hierarchy it was  $-8\lesssim \log_{10}\lambda_{\phi s}\lesssim -3$). Accordingly, smaller values of $\lambda_{\phi s}$ mean that fewer mediators annihilate to DM and less heat is transferred to it from the mediator fluid, thereby necessitating larger FI production (since DM lacks sufficient kinetic energy to increase its number density via cannibal processes), which exacerbates the constraints. Consequently, larger values of $\lambda_{\phi s}$ are preferred.

\begin{figure}[t!]
    \centering
    \begin{subfigure}
        \centering
        \includegraphics[width=0.48\textwidth]{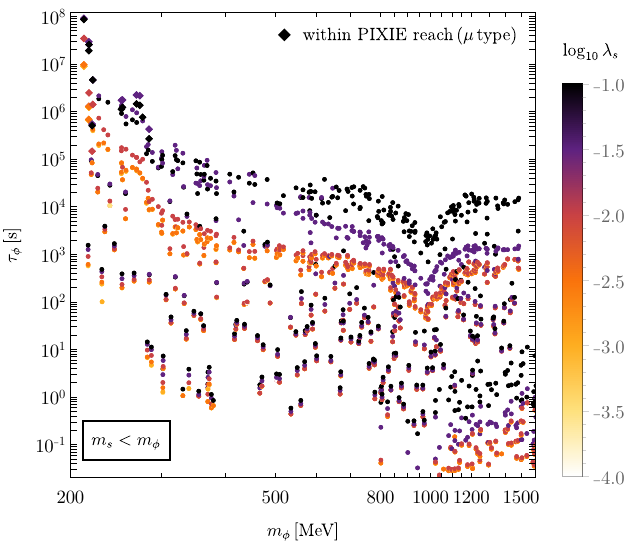}
    \end{subfigure}
    \hfill
    \begin{subfigure}
        \centering
        \includegraphics[width=0.48\textwidth]{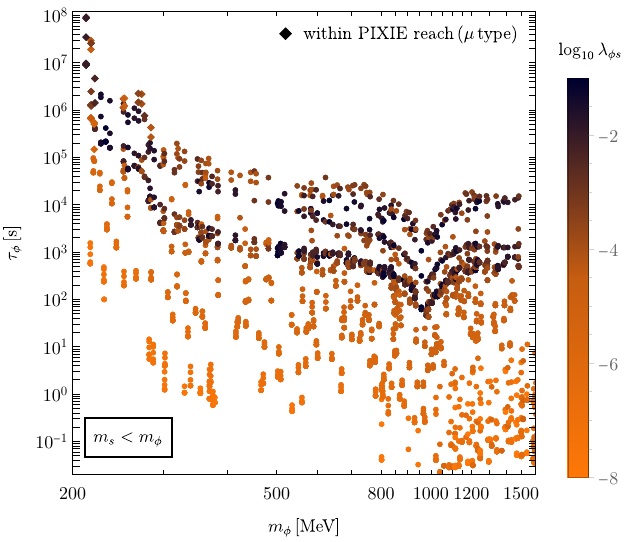}
    \end{subfigure}
    \caption{Same as Figure~\ref{fig:tauphi_vs_abundance_ms<mphi} for the case $m_s<m_\phi$ with solutions above the muon threshold. Here potential testable points are found for $\phi$ masses between $212$ and $284.35\,\text{MeV}$, with $\log_{10}\lambda_{\phi s}\in[-5.27,-2]$.}
    \label{fig:tauphi_vs_abundance_ms<mphi_GeV}
\end{figure}

\begin{figure}[t!]
     \centering
     \begin{subfigure}
        \centering
        \includegraphics[width=0.48\textwidth]{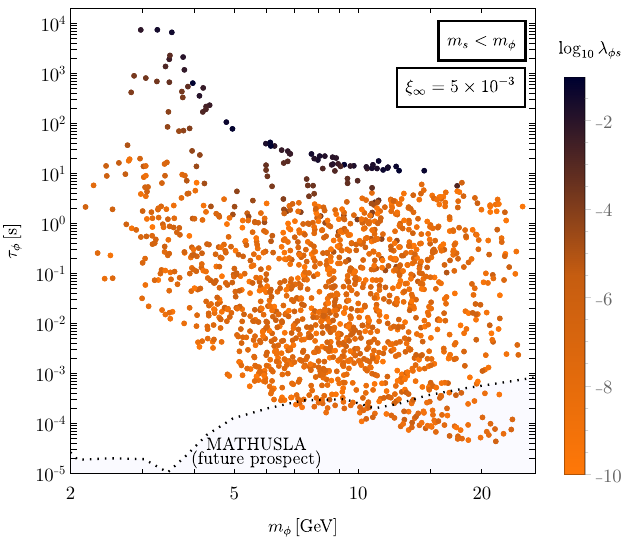}
    \end{subfigure}
    \hfill
    \begin{subfigure}
        \centering
        \includegraphics[width=0.49\textwidth]{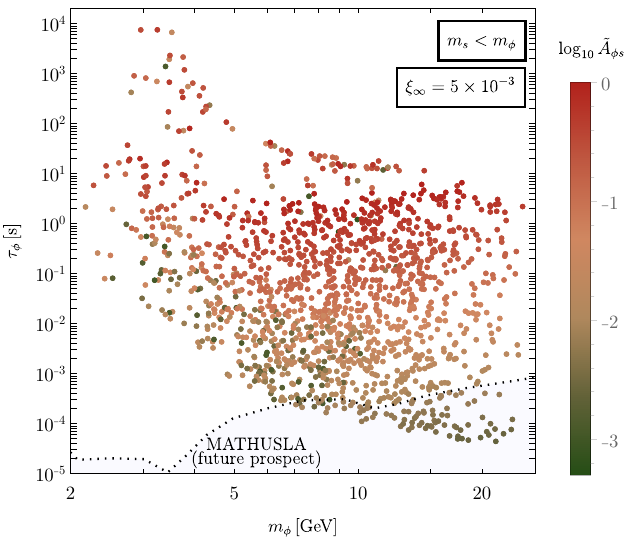}
    \end{subfigure}
    \caption{Results in the GeV range with the superimposed prospect for the future experiment MATHUSLA~\cite{Curtin:2024xxo}. Potential detectable points satisfy $\log_{10}\lambda_{\phi s}\lesssim-6.15$, where the mediator and DM sectors are disconnected and both are produced independently.
    }
    \label{fig:Mathusla}
\end{figure}

The influence of parameter \(\lambda_{s}\)  is firstly seen through apparent discrete bands the points are align into, most clearly observed in Figure~\ref{fig:tauphi_vs_abundance_ms<mphi_GeV}, but discernible in other plots as well. This effect does not carry any significant meaning, as it is a consequence of a discrete grid in $\lambda_s$ sampling. In particular, the visible bands correspond to \(\log_{10}\lambda_s = -2, -1.5\), and \(-1\). What is however physical, is that the lower $\lambda_s$ is, the shorter the mediator lifetime, as weaker self-interactions mean weaker boosting of the FI production, thus increasing the mediator coupling (cf. \Cref{fig:evol1,fig:evol3}).
Additionally, the parameter \(\lambda_{\phi s}\) also influences the distribution of points, segregating into two regimes: \(\log_{10}\lambda_{\phi s} \lesssim -2.5\) and \(\log_{10}\lambda_{\phi s} \gtrsim -2.5\) (Figure~\ref{fig:tauphi_vs_abundance_ms<mphi_GeV} right). Note that the distinctive resonance-like behavior observed at $m_\phi = 1\,\text{GeV}$ is primarily a consequence of the sudden decrease in the $\phi$ lifetime due to a resonance in the $\Gamma_{\phi\to\pi\pi }$ rate~\cite{Winkler:2018qyg}.

Last but not least, the $m_s<m_\phi$ case can also be tested in the Massive Timing Hodoscope for Ultra-Stable Neutral Particles (MATHUSLA) experiment designed to detect exotic long-lived particles generated by LHC collisions. These particles would purportedly be capable of traveling to the surface of the collider's detector, where they may decay into SM charged particles. The detector will possess a sensitivity to lifetimes as large as $10^{-4}\,\text{s}$~\cite{Curtin:2024xxo}, rendering GeV mediators in this model to be within reach. The results are shown in Figure~\ref{fig:Mathusla}, where we highlight the role of $\lambda_{\phi s}$ and $\tilde A_{\phi s}(=A_{\phi s}/m_\phi)$. Testable points satisfy $\log_{10}\lambda_{\phi s}\lesssim -6.15$ and $-2.9\lesssim\log_{10}\tilde A_{\phi s}\lesssim -2$, that is, both sectors are essentially decoupled and are populated separately: $\phi$ directly through $\theta$, while $S$ via Higgs decay $A_{\phi s}\theta\,h\vert S\vert^2$. This result can be understood as follows. The hierarchy $m_s<m_\phi$ pushes the mediator mass towards the GeV scale and small $\lambda_{\phi s}$ ensures that DM is not produced from mediator annihilation, implying a larger $\theta$ than otherwise, which shortens $\tau_\phi$. Similarly, a larger $\tilde A_{\phi s}$ would lead to DM overproduction from Higgs decay. This can be compensated with a lower $\theta$, resulting in longer lifetimes beyond the detector's range.

In summary, the mass hierarchy $m_s<m_\phi$ opens up the parameter space in the MeV range and moreover adds further motivation to searches planned to be conducted in MATHUSLA and PIXIE.

\clearpage
\section{Conclusions}
\label{sec:conclusions}

Cannibalization is an intriguing mechanism for depleting the number of dark matter particles without the necessity of coupling the dark sector to the SM plasma. As such, it provides an alternative to the standard thermal freeze-out, featuring a different phenomenological profile and complex decoupling dynamics. For it to be successful, however, a large initial population of dark sector states must first be produced, which subsequently undergo a cannibalization phase. It is also crucial to ensure that heat released in this process does not interfere with the structure formation. Both conditions can be satisfied by introducing a very small coupling between the dark sector and the SM, which facilitates freeze-in type production of a dark sector with an initial temperature significantly lower than that of the SM plasma. 

In this work, we investigate three such scenarios with a particular emphasis on whether the coupling strength required for effective dark matter production can also yield detectable signals in cosmological probes, indirect detection, and searches for long-lived particles. In all cases we derive and solve the coupled Boltzmann equations for the number density and temperature for all particles taking part in the freeze-in and freeze-out processes. Implementation of all the relevant reactions, i.e., decays, annihilations, elastic scatterings, and $3\leftrightarrow 2$ interactions, allows us to accurately analyze the interplay of the heat transfer between the visible and dark sectors and the evolution of the number densities of all the states.

The first model we study consists of the dark matter being a self-interacting real scalar field in a $\mathbb{Z}_2$ broken phase. Spontaneous breaking of the stabilizing symmetry provides both the $3\leftrightarrow 2$ reactions and induces DM decay. The latter leads to stringent constraints on the coupling to the SM, but also to a potential for detection of otherwise a completely secluded model. The results shown in Figure~\ref{fig:resultsZ2} indicate that indeed the parameter space of such a model is tightly constrained by several factors: observations of the Bullet Cluster, constraints from the INTEGRAL and NuSTAR on decaying dark matter, and the requirement that the dark matter lifetime is larger than the age of the universe. The portal coupling allowed by these observations is found to be small enough that the accompanying freeze-in mechanism is not strong enough to  populate the dark sector completely and such a scenario is shown to require an additional production mechanism.

In the second model we explore a dark sector containing still only one state, now a complex dark matter candidate $S$ stabilized by $\mathbb{Z}_3$ symmetry. We show that the freeze-in mechanism can successfully account for the observed relic abundance (cf. Figure~\ref{fig:resultsZ3}) while not contradicting known observations. During the analysis we highlight the non-trivial dynamics of boosting the freeze-in production by $2\to 3$ self-interactions. 
Thus we show that a $\mathbb{Z}_3$ scalar singlet model possesses a valid alternative to its freeze-out and freeze-in production, with the intermediate cannibal phase. There is, however, no current technology to test this scenario, as in contrast to the previous case there are no DM decays present and all interactions with the SM are mediated by the Higgs boson with a very small mixing to $S$.

The third scenario extends the previous model by including an unstable Higgs-like real mediator $\phi$. In this setup we account for the DM-mediator interactions by solving the coupled Boltzmann equations for the system of two dark sector particles and performing a numerical parameter scan over DM self-interactions, as well as DM-mediator couplings. We identify two scenarios with distinct phenomenology. In the first case, $m_s>m_\phi$, the kinematically allowed  $SS^*\to2\phi$ process typically results in an overabundance of mediators during the BBN epoch, leading to stronger constraints (Figure~\ref{fig:tauphi_vs_abundance_ms>mphi}). Therefore, both dark matter self- and DM-mediator interactions turn out to be crucial in mitigating cosmological constraints on the mediator's lifetime. For example, feeble $S-\phi$ interactions result in a shorter lifetime (Figure~\ref{fig:tauphi_vs_abundance_ms>mphiGeV}), allowing the model to evade BBN constraints. Indirect detection signatures are also feasible because the s-wave annihilation cross section can be sizable (Figure~\ref{fig:cross_ID}). Conversely, if $m_s<m_\phi$, mediators are depleted to produce more of dark matter during the evolution in the early universe, thereby relaxing cosmological bounds on their energy injection to the SM plasma. Specifically, with the dark matter and mediator at the GeV range there is a potential for detectable signals in long-lived particle searches. In particular projections for MATHUSLA are depicted in Figure~\ref{fig:Mathusla}.

To summarize, our study demonstrates that a frozen-in dark sector scenario featuring a cannibal dark matter candidate can simultaneously satisfy the observed abundance of DM, adhere to known constraints  and hold some promise for detection in upcoming experiments like MATHUSLA or leave distortions in the CMB that could be in the sensitivity range of PIXIE. Moreover, we highlight that interactions contained solely within the dark sector can significantly influence the signatures in these experiments.

\acknowledgments
It is a pleasure to thank Shiuli Chatterjee for insightful discussions. This work was supported by the National Science Centre (Poland) under the research Grant No. 2021/42/E/ST2/00009.

\appendix
\section{Collision operators for number changing self-interactions}\label{ap:A}
\subsection{Matrix elements}\label{ap:Matrix_elem_self}
The matrix element for the reaction $\varphi_1\varphi_2\leftrightarrow\varphi_3\varphi_4\varphi_5$ corresponding to the real scalar model is 
\begin{equation}
\begin{split}
        \label{eq:mat_el_real}
        i\mathcal{M}_{3\varphi\leftrightarrow 2\varphi} = -i g^3\Bigg(&
            \frac{1}{S S_{34}} +\frac{1}{S S_{35}}+\frac{1}{S S_{45}} +\frac{1}{T_{15}S_{34}} +\frac{1}{T_{14}S_{35}}+\frac{1}{T_{13}S_{45}} 
            \\&+\frac{1}{T_{25}S_{34}} +\frac{1}{T_{24}S_{35}}+\frac{1}{T_{23}S_{45}} 
            +\frac{1}{T_{14}T_{23}} +\frac{1}{T_{15}T_{23}}+\frac{1}{T_{13}T_{24}} 
            \\&\hspace{13em}+\frac{1}{T_{14}T_{25}} +\frac{1}{T_{13}T_{25}}+\frac{1}{T_{15}T_{24}}\Bigg) 
        \\&\hspace{-4.4em}-ig\lambda\Bigg(\frac{1}{S}+\frac{1}{S_{34}} +\frac{1}{S_{35}}+\frac{1}{S_{45}}+\frac{1}{T_{13}}+\frac{1}{T_{14}}
        +\frac{1}{T_{15}}+\frac{1}{T_{23}} +\frac{1}{T_{24}}+\frac{1}{T_{25}}\Bigg)\,,
\end{split}
\end{equation}
where we have defined $S_{ij}=s_{ij}-m_\varphi^2$, $T_{ij} = t_{ij}-m_\varphi^2$, with $S=S_{12}$. In the broken phase $g=\sqrt{3\lambda}m_\varphi$. This means that we can factorize $\sqrt{3\lambda^3}$, along with the mass in the propagator by defining the dimensionless Mandelstam variables $\tilde s_{ij}= s_{ij}/m_\varphi^2$ and $\tilde t_{ij}=t_{ij}/m_\varphi^2$. The factor is $\sqrt{3\lambda^3}/m_\varphi$. 

In the complex scalar model we encounter two matrix elements, the first one is denoted as $\mathcal{M}_{S^*S\leftrightarrow SSS}$ (short $\mathcal{M}^{(1)}$), the result is
\begin{equation}
    \begin{split}
        i\mathcal{M}^{(1)}=&-ig^2\left(\frac{1}{S_{45}T_{13}}+\frac{1}{S_{35}T_{14}}+\frac{1}{S_{34}T_{15}} \right)
        \\&-ig\lambda_s\left(\frac{1}{S_{34}}+\frac{1}{S_{35}}+\frac{1}{S_{45}}+\frac{1}{T_{13}}+\frac{1}{T_{14}}+\frac{1}{T_{15}}\right)\,.
    \end{split}
\end{equation}
In this case $S_{ij} = s_{ij}-m_s^2$, analogously defined for $T_{ij}$. Finally, the matrix element corresponding to $SS\leftrightarrow S^*S^*S$ (denoted as  $\mathcal{M}^{(2)}$) is 
\begin{equation}
    \begin{split}
        i\mathcal{M}^{(2)}=&-ig^2\left(\frac{1}{SS_{34}}+\frac{1}{T_{14}T_{23}}+\frac{1}{T_{13}T_{24}}  \right)
        \\&-ig\lambda_s\left(\frac{1}{S}+\frac{1}{S_{34}}+\frac{1}{T_{13}}+\frac{1}{T_{14}}+\frac{1}{T_{23}}+\frac{1}{T_{24}}\right)\,.
    \end{split}
\end{equation}

\subsection{$3\varphi\leftrightarrow 2\varphi$ collision integrals}\label{ap:C0self}
The zeroth moment collision term is
\begin{equation}
    \begin{split}
        &\braket{C_{3\varphi\leftrightarrow 2\varphi}} = \frac{g_\varphi}{n_\varphi}\int\frac{d^3\vec p}{(2\pi)^3} C_{3\varphi\leftrightarrow 2\varphi} 
        \\&= \frac{1}{ n_\varphi }\frac{1}{2!}\frac{1}{3!}
        \int d\Pi_1 \dots d\Pi_5\,\vert\tilde{\mathcal{M}}_{3\varphi\leftrightarrow 2\varphi}\vert^2\left(f_1f_2-f_3f_4f_5\right)
        \\&=\frac{1}{12 n_\varphi} \,\int d\Pi_1 \dots d\Pi_5\,\vert\tilde{\mathcal{M}}_{3\varphi\leftrightarrow2\varphi}\vert^2 e^{-(E_1+E_2)/T_\varphi}\left( \left(\frac{n_\varphi}{n^\text{eq}_\varphi} \right)^2 - \left(\frac{n_\varphi}{n^\text{eq}_\varphi} \right)^3 \right)
         \\&=s Y_\varphi\braket{\sigma_{3\varphi\to 2\varphi}v^2}s\,(Y_\varphi^\text{eq} - Y_\varphi)\,,
    \end{split}
    \label{eq:C0_self}
\end{equation}
where $g_\varphi$ accounts for the DM degrees of freedom ($g_\varphi = 1$). Note also that we used $f_3 f_4 f_5 = \frac{n_\varphi}{n^\text{eq}_\varphi}f_1 f_2$. 

The three final state integral can be evaluated by first boosting to the lab frame (CM of two final states, e.g. $\varphi_3\varphi_4$), followed by a boost to the total CM frame~\cite{Byckling:1971vca}. In the lab frame $p_1+p_2-p_5=(\sqrt{s_{34}},\vec 0\,)^\intercal$, with $s_{34}=(p_3+p_4)^2 = (p_1+p_2-p_5)^2 = s+m_\varphi^2-2\sqrt{s}E_5$. 
By definition $p_3^\text{lab} = (\sqrt{s_{34}}/2,\vec p_3^\text{ lab})^\intercal$ and $p_4^\text{lab} = (\sqrt{s_{34}}/2,-\vec p_3^\text{ lab})^\intercal$. We can relate the lab with the CM frame via a boost in the $z$ direction followed by a rotation around the $y$ axis,
\begin{equation}
    \mathcal{B}_z(\gamma) = \begin{pmatrix}
        \gamma & 0 & 0 &\gamma\beta
        \\0 & 1 & 0 & 0
        \\ 0 & 0 & 1 & 0
        \\ \gamma\beta & 0 & 0 & \gamma
    \end{pmatrix}\,,\qquad \mathcal{R}_y(\alpha) = \begin{pmatrix}
        1 & 0 & 0 & 0
        \\ 0 & \cos\alpha & 0 & \sin\alpha
        \\ 0 & 0 & 1 & 0
        \\ 0 &-\sin\alpha & 0 &\cos\alpha
    \end{pmatrix}\,,
\end{equation}
such that $p_i^\text{cm} = \mathcal{R}_y(\alpha+\pi)\mathcal{B}_z(\frac{E_{12}}{\sqrt{s_{34}}}) p_i^\text{lab}$, where $E_{12} = (s + s_{34} - m_\varphi^2)/(2\sqrt{s})$, and $\beta=\sqrt{1-1/\gamma^2}$. After applying these transformations, the thermal average in the CM frame is
\begin{equation}\label{eq:3to2_average}
   \begin{split}
   &\braket{\sigma_{3\varphi\to 2\varphi}v^2}=\frac{1}{2!3!}\frac{1}{(n^\text{eq}_\varphi)^3}\int d\Pi_1 \dots d\Pi_5\,\vert\tilde{\mathcal{M}}_{3\varphi\leftrightarrow2\varphi}\vert^2 e^{-(E_1+E_2)/T_\varphi}
   \\&=\frac{1}{12}\frac{1}{(n_\varphi^\text{eq})^3}\frac{3}{8(2\pi)^4}\int d\Pi_1 d\Pi_2 e^{-\frac{E_1+E_2}{T_\varphi}}\int d\tilde E_5\, d\phi_4\,dx_4\,dx_5\,J \,\vert\mathcal{M}(\tilde s_{ij},\tilde t_{ij})\vert^2  \,,
   \end{split}
\end{equation}
where $\tilde E_5 = E_5/m_\varphi$ and the limits of integration are $ \tilde E_5\in [1,(s-3m_\varphi^2)/(2m_\varphi \sqrt{s})]$, $x_{4,5}(=\cos\theta_{4,5})\in[-1,1]$ and $\phi_4\in[0,2\pi)$. Additionally, 
\begin{equation}
    J=\sqrt{\tilde E_5^2 -1}\sqrt{1-4m_\varphi^2/(m_\varphi^2-2\tilde E_5 m_\varphi\sqrt{s}+s)}\,. 
\end{equation}
We evaluate the 4-d integral numerically with the Monte-Carlo method. The integral over the initial states can be evaluated in terms of $E_+=E_1+E_2$ and $s$~\cite{Edsjo:1997bg},
\begin{equation}
d\Pi_1 d\Pi_2 = \frac{1}{(2\pi)^4}\frac{1}{8}dE_+dE_-ds = \frac{1}{(2\pi)^4}\frac{p_{12}}{2}\sqrt{\frac{E_+^2-s}{s}}dE_+ ds\,,
\end{equation}
with $p_{12} = \frac{1}{2}\sqrt{s-4m_\varphi^2}$. Finally, we note that the thermal average scales inversely with the mass to the fifth power in the non-relativistic limit, 
\begin{equation}\label{eq:ap_m_scaling}
    \braket{\sigma_{3\varphi\to 2\varphi }v^2}\sim \lambda^3\,m_\varphi^4/(n_\varphi^\text{eq})^3\sim  \lambda^3/m_\varphi^5\,.
\end{equation} 
The second moment collision term is 
\begin{equation}
    \begin{split}
    3T_\varphi n_\varphi \braket{C_{3\varphi\leftrightarrow 2\varphi}}_2 = &-\frac{1}{3!}\int d\Pi_1\dots d\Pi_5
        \frac{\vec p_1^{\,2}}{E_1} f_1 f_2\vert\tilde{\mathcal{M}}_{3\varphi\leftrightarrow 2\varphi}\vert^2
        \\&+\frac{1}{2!2!}\int d\Pi_1\dots d\Pi_5 \frac{\vec p_3^{\,2}}{E_3} f_1 f_2 \vert\tilde{\mathcal{M}}_{3\varphi\leftrightarrow 2\varphi}\vert^2
        \\&-\frac{1}{2!2!}\int d\Pi_1\dots d\Pi_5 \frac{\vec p_3^{\,2}}{E_3} f_3 f_4 f_5 \vert\tilde{\mathcal{M}}_{3\varphi\leftrightarrow 2\varphi}\vert^2
        \\&+\frac{1}{3!}\int d\Pi_1\dots d\Pi_5 \frac{\vec p_1^{\,2}}{E_1} f_3 f_4 f_5 \vert\tilde{\mathcal{M}}_{3\varphi\leftrightarrow 2\varphi}\vert^2\,.
    \end{split}
    \label{eq:C2_self}
\end{equation}
The terms with $1/3!$ combine to
\begin{equation}
    \braket{\sigma_{3\varphi\to 2\varphi} v^2\,\vec p^{\,2}/E }n_\varphi^2\,(n_\varphi-n_\varphi^\text{eq})\,,
\end{equation}
where
\begin{equation}
   \braket{\sigma_{3\varphi\to 2\varphi} v^2\,\vec p^{\,2}/E }=\frac{g_\varphi^3}{(n_\varphi^\text{eq})^3}\frac{1}{3!}\int d\Pi_1 d\Pi_2 \frac{\vec p_1^{\,2}}{E_1}f_1f_2\,4F\,\sigma_{2_\varphi\to3_\varphi}\,,
\end{equation}
and $F=\sqrt{\left(s/2-m_\varphi^2\right)^2-m_\varphi^4}$. On the other hand, the terms with factor $1/(2!2!)$ lead to
\begin{equation}
\frac{1}{2!2!}\left(\frac{n_\varphi}{n_\varphi^\text{eq}} \right)^2\int d\Pi_1 \dots d\Pi_5\,f_1f_2\,\frac{\vec p_3^{\,2}}{E_3}\vert\tilde{\mathcal{M}}_{3\varphi\leftrightarrow2\varphi}\vert^2\left( 1-\frac{n_\varphi}{n^\text{eq}_\varphi}\right)\,.
\label{eq:C2_2}
\end{equation}
The three final states integral is $\int d\Pi_3 d\Pi_4 d\Pi_5 (\vec p_3^{\,2}/E_3)\vert\tilde{\mathcal{M}}_{3\leftrightarrow2}\vert^2 $, which is not Lorentz invariant due to the factor $\vec p_3^{\,2}/E_3$. Boosting the integrand to the CM of mass frame, the energy transforms as~\cite{Arcadi:2019oxh}
\begin{equation}\label{eq:Energy_to_CM}
    E_3\to E_3\,\cosh{\eta}+\vec p^{\,z}_3\,\sinh\eta\,,\qquad\text{with}\qquad \vec p^{\,z}_3 = \sqrt{E_3^2-m_\varphi^2}\cos\theta_{3}\,.
\end{equation}
In this case $\eta$ is the rapidity. This suggests to define
\begin{equation}\label{eq:sigma_tilde}
\begin{split}
   4F\tilde\sigma_{2\varphi\to 3\varphi}(\eta)=&\frac{1}{2!2!}\int d\Pi_3 d\Pi_4 d\Pi_5\,(\vec p_3^{\,2}/E_3)\vert\tilde{\mathcal{M}}_{3\varphi\leftrightarrow2\varphi}\vert^2
\end{split}
\end{equation}
and its thermal average as
\begin{equation}
\begin{split}
    \braket{\tilde \sigma_{2\varphi\to 3\varphi} v }=\frac{g_\varphi^2}{(2\pi)^4 (n_\varphi^\text{eq})^2}&\int_{3m_\varphi}^\infty dE_\text{cm}\sqrt{\frac{E_\text{cm}^2}{4}-m_\varphi^2}E_\text{cm}^2
    \\&\times\int_0^\infty d\eta \sinh^2\eta\,\exp\left({-\frac{E_\text{cm}}{T_\varphi}\cosh\eta}\right)\,4F\tilde\sigma_{2\varphi\to 3\varphi}(\eta)\,,
\end{split}
\end{equation}
then, the second moment, $\braket{C_{3\varphi\leftrightarrow 2\varphi}}_2$, in terms of the comoving number density takes the form
\begin{equation}
\begin{split}\label{eq:C2_real_cannibal_scalar}
    3T_\varphi \braket{C_{3\varphi\leftrightarrow 2}}_2
    =s^2Y_\varphi(Y_\varphi-Y_\varphi^\text{eq})\left(\braket{\sigma_{3\varphi\to 2\varphi} v^2\,\vec p^{\,2}/E} -\braket{\tilde \sigma_{3\varphi\to 2\varphi}v^2}\right)\,.
\end{split}
\end{equation}

\subsection{$3\leftrightarrow 2$ collision integrals for complex DM}\label{ap:C_Z3}
%
Considering the reaction $SS\leftrightarrow S^*S^* S$, the collision integral for $S^*$ is
\begin{equation}
    \begin{split}
        C_{SS\leftrightarrow S^*S^*S}[S^*]=\frac{1}{2E_{S^*}g_{S^*}}&\int\Bigg(
        (1+f_{S^*})\vert\tilde{\mathcal{M}}_{12\to \underline{S^*}45}\vert^2\left(\frac{1}{2!}d\Pi_1d\Pi_2f_1f_2\right)d\tilde\Pi_4d\tilde\Pi_5
        \\&-f_{S^*}\vert\tilde{\mathcal{M}}_{\underline{S^*}45\to 12}\vert^2 d\Pi_4d\Pi_5 f_4 f_5 \left(\frac{1}{2!}d\tilde\Pi_1d\tilde\Pi_2 \right)
        \Bigg)\,,
    \end{split}
\end{equation}
while for $S$ is
\begin{equation}
    \begin{split}
        C_{SS\leftrightarrow S^*S^*S}[S] =\frac{1}{2E_{S}g_{S}} &\int\Bigg( 
        -f_S\vert\tilde{\mathcal{M}}_{\underline{S}2\to 345}\vert^2 d\Pi_2f_2\left(\frac{1}{2!}d\tilde\Pi_3d\tilde\Pi_4d\tilde\Pi_5\right)
        \\&+(1+f_S)\vert\tilde{\mathcal{M}}_{12\to 34\underline{S}}\vert^2\left(\frac{1}{2!}d\Pi_1d\Pi_2 f_1f_2\right)\left(\frac{1}{2!}d\tilde\Pi_3d\tilde\Pi_4\right)
        \\&-f_S\vert\tilde{\mathcal{M}}_{34\underline{S}\to12}\vert^2\left(\frac{1}{2!}d\Pi_3d\Pi_4f_3f_4\right)\left(\frac{1}{2!}d\tilde\Pi_1d\tilde\Pi_2\right)
        \\&+(1+f_S)\vert\tilde{\mathcal{M}}_{345\to \underline{S}2}\vert^2\left(\frac{1}{2!}d\Pi_3d\Pi_4d\Pi_5f_3f_4f_5\right)d\tilde\Pi_2
        \Bigg)\,.
    \end{split}
\end{equation}
As we assume no CP violation, $\mathcal{M}_{S^*S\leftrightarrow SSS} =\mathcal{M}_{SS^*\leftrightarrow S^*S^*S^*}$ and $\mathcal{M}_{SS\leftrightarrow S^*S^*S} = \mathcal{M}_{S^*S^*\leftrightarrow SSS^*}$, we denote $\mathcal{M}^{(1)}$ the former and $\mathcal{M}^{(2)}$ the latter. Since we assume $f_S=f_{S^*}$ then $n_{S} = n_{S^*}$. Notice also that the total abundance is $n=n_S+n_{S^*}$. Neglecting Bose-Enhancement terms (and setting $g_S=1$), 
\begin{equation}
    \int \frac{d^3\vec p_{s^*}}{(2\pi)^3}C[S^*] = \frac{1}{2}\int d\Pi_1\dots d\Pi_5\,\left(\frac{1}{3}\vert\tilde{\mathcal{M}}^{(1)}\vert^2 + \frac{1}{2}\vert\tilde{\mathcal{M}}^{(2)}\vert^2\right)\,(f_1 f_2-f_3 f_4 f_5)\,.
\end{equation}
Note that $\int \frac{d^3\vec p_s}{(2\pi)^3}\,C[S]$ retains the same expression. The second moment collision integral is
\begin{equation}
    \begin{split}
        \int \frac{d^3\vec p_1}{(2\pi)^3} \frac{\vec p^{\,2}_1}{E_1}C[S^*] = &-\frac{1}{3}\int d\Pi_1\dots d\Pi_5\vert\tilde{\mathcal{M}}^{(1)}\vert^2\frac{\vec p_1^{\,2}}{E_1}(f_1f_2-f_3f_4f_5) 
        \\&+\frac{1}{2}\int d\Pi_1\dots d\Pi_5 \vert\tilde{\mathcal{M}}^{(1)}\vert^2\frac{\vec p^{\,2}_3}{E_3}(f_1f_2-f_3f_4f_5)
        \\&-\frac{1}{2}\int d\Pi_1\dots d\Pi_5\vert\tilde{\mathcal{M}}^{(2)}\vert^2\frac{\vec p_1^{\,2}}{E_1}(f_1f_2-f_3f_4f_5)
        \\&+\frac{3}{4}\int d\Pi_1\dots d\Pi_5\vert\tilde{\mathcal{M}}^{(2)}\vert^2\frac{\vec p_3^{\,2}}{E_3}(f_1f_2-f_3f_4f_5)\,,
    \end{split}
\end{equation}
which results in
\begin{equation}
    \begin{split}
     \frac{3}{2}\int &d\Pi_1\dots d\Pi_5\vert\tilde{\mathcal{M}}\vert^2 \frac{\vec p^{\,2}_3}{E_3}(f_1f_2-f_3f_4f_5) 
     \\&- \int d\Pi_1 \dots d\Pi_5\vert\tilde{\mathcal{M}}\vert^2\frac{\vec p_1^{\,2}}{E_1}(f_1f_2-f_3f_4f_5)\,,
    \end{split}
\end{equation}
with $\vert\tilde{\mathcal{M}}\vert^2=\frac{1}{3}\vert\tilde{\mathcal{M}}^{(1)}\vert^2 + \frac{1}{2}\vert\tilde{\mathcal{M}}^{(2)}\vert^2$. Similar to Section~\ref{ap:C0self}, the thermal averages in Eq.~\eqref{eq:C2_real_cannibal_scalar} remain the same, with the only difference being the replacement of the corresponding matrix element and Boltzmann factors. 
%
\section{Freeze-in collision integrals}\label{ap:CFI}
\subsection{Higgs decay}
The zeroth moment collision integral for Higgs decay is
\begin{equation}
    n_\chi\braket{C_{h\leftrightarrow\chi\chi}} = \int d\Pi_1d\Pi_2 d\Pi_3\vert \tilde{\mathcal{M}}_{h\leftrightarrow\chi \chi}\vert^2
    \left(f_1-f_2f_3\right)\,,
\end{equation}
where $\chi$ stands for either $\varphi$ or $S$. We have labeled the momenta as $h_1\leftrightarrow \chi_2\chi_3$. Since $\mathcal{M}$ is constant in this case, we can pull it out of the integration. Neglecting Higgs production,
\begin{equation}\label{eq:ap_Higgs_decay_C0}
\begin{split}
    n_\chi\braket{C_{h\to\chi\chi}}&=\frac{c^2}{16\pi^2}\int d\Pi_1 f_1 \int \frac{d^3\vec p_2}{E_2}
   \frac{d^3\vec p_3}{E_3}
   \delta^{(4)}(p_1-p_2-p_3)
   \\&=c^2\frac{m_h}{16\pi^3}\sqrt{1-\frac{4m_\chi^2}{m_h^2}} T K_1(m_h/T)\,,
\end{split}
\end{equation}
where $c$ stands for the coupling constant in question. For interactions with $\varphi$ it is $c = \lambda_{h\varphi} v$, while for $S$ it is $c=A_{\phi s}\theta$. Notice the difference of a factor 2 due to different conventions~\cite{Lebedev:2019ton}, stemming from the definition of symmetry factors. 

On the other hand, the second moment is
\begin{equation}
    3T_\chi n_\chi \braket{C_{h\to \chi\chi}}_2 = \int d\Pi_1d\Pi_2 d\Pi_3\, \vert \tilde{\mathcal{M}}_{h\to\chi\chi}\vert^2 \frac{\vec p_2^{\,2}}{E_2}f_1\,.
\end{equation}
The term $\vec p_2^{\,2}/E_2$ is not Lorentz invariant, after boosting to the CM frame,
\begin{equation}
   E_2 \to E_2\,\cosh\eta+\vec p_2^{\,z}\sinh\eta\,,
\end{equation}
which we cast in terms of $z=\cosh\eta$. Note that $z=E_h/m_h$.
\begin{equation}\label{eq:ap_CMboost}
\frac{\vec p_2^{\,2}}{E_2} \to E_2 z + \vert \vec p_2\vert \cos\theta_2 \sqrt{z^2-1} - \frac{m_\chi^2}{E_2 z + \vert \vec p_2\vert \cos\theta_2 \sqrt{z^2-1}}\,.
\end{equation}
Thus, $3T_\chi n_\chi\braket{C_{h\to\chi\chi}}_2$ results in
\begin{equation}\label{eq:ap_Higgs_decay_C2}
\begin{split}
     &\frac{c^2}{4(2\pi)^2}\int d\Pi_1f_1\int d\cos\theta_2\,\frac{2\pi \vert \vec p_2\vert}{E_2}\left.\left(E_2\,z - \frac{m_\chi^2}{E_2 z + \vert \vec p_2\vert \cos\theta_2 \sqrt{z^2-1}}\right)\right\vert_{E_2=m_h/2}
    \\&=\frac{c^2 m_h^2}{32\pi^3}\sqrt{1-\frac{4m_\chi^2}{m_h^2}}\Bigg(T K_2(m_h/T)
    \\&\hspace{8em}-\frac{2m_\chi^2}{\sqrt{m_h^2-4m_\chi^2}}\int_1^\infty dz\,e^{-m_h z/T}\log\left(\frac{m_h z + 2\vert\vec p_2\vert\sqrt{z^2-1}}{m_h z - 2\vert\vec p_2\vert\sqrt{z^2-1}}\Bigg)\right)\,.
\end{split}
\end{equation}
If there is a clear mass hierarchy, $m_\chi\ll m_h$, the logarithmic term can be safely ignored.

\subsection{Three-body Higgs decay}\label{ap:triple_higgs_decay}
The zeroth moment for the mediator $\phi$ is
\begin{equation}
    g_\phi n_\phi \braket{C_{h\leftrightarrow \phi S S^* }} = \int d\Pi_1d\Pi_2 d\Pi_3 d\Pi_4\vert \tilde{\mathcal{M}}_{h\leftrightarrow \phi S S^* }\vert^2
    \left(f_1-f_2f_3f_4\right)\,,
\end{equation}
where the momenta are labeled as $h_1\leftrightarrow \phi_2S_3S^*_4$. As in the previous case, we neglect Higgs production, 
\begin{equation}
    n_\phi \braket{C_{h\to \phi S S^* }} = \frac{\lambda_{\phi s}^2\theta^2}{256\pi^5}\int d\Pi_1\,f_1 \int \frac{d^3\vec p_2}{E_2}
   \frac{d^3\vec p_3}{E_3} \frac{d^3\vec p_4}{E_4}
   \delta^{(4)}(p_1-p_2-p_3-p_4)\,.
\end{equation}
The final three state integral can be evaluated as in Section.~\ref{ap:C0self} with $s_{34}=(p_3+p_4)^2 = (p_1-p_2)^2 = m_h^2+m_\phi^2-2m_hE_2$, the result is
\begin{equation}
   8\pi^2  \int_{m_\phi}^{E_2^\text{max}} dE_2\, \vert \vec p_2\vert^{2} \sqrt{1-\frac{4m_s^2}{m_h^2+m_\phi^2-2m_hE_2}}\,,
\end{equation}
where 
\begin{equation}
\begin{split}
     E_2^\text{max} &= \sqrt{(\vec q_2^\text{ max})^2+m_\phi^2}\qquad \text{and} 
     \\\vert \vec q_2^\text{ max}\vert &= \sqrt{m_h^4+(m_\phi^2-4m_s^2)^2-2m_h^2(m_\phi^2+4m_s^2)}/(2m_h)\,,
\end{split}
\end{equation}
with $\vert \vec p_2^\text{ max}\vert$ the maximum magnitude of the momentum allowed for $\phi$. The previous integral can be evaluated analytically in the limit $m_\phi,\,m_s\ll m_h$. The result is $\pi^2 m_h^2$. Thus, the zeroth moment is
\begin{equation}
\begin{split}
    n_\phi \braket{C_{h\to \phi SS^*}}&\simeq  \frac{\lambda_{\phi s}^2\theta^2}{1024\pi^5}\,m_h^4 \int_1^\infty dz\,\sqrt{z^2-1}\exp(-m_h z/T)
     \\&=\frac{\lambda_{\phi s}^2\theta^2}{1024\pi^5}m_h^3 T K_1(m_h/T)\,. 
\end{split}
\end{equation}

Finally, the second moment is
\begin{equation}
    3T_\phi n_\phi \braket{C_{h\to \phi S S^* }}_2 = \int d\Pi_1d\Pi_2 d\Pi_3 d\Pi_4\vert \tilde{\mathcal{M}}_{h\to \phi S S^* }\vert^2\,\frac{\vec p_2^{\,2}}{E_2}\,f_1\,.
\end{equation}
Analogous to the double Higgs decay, we boost $\vec p_2^{\,2}/E_2$ to the CM frame. Here we already adopt the massless limit and neglect the logarithmic contribution. The result is 
\begin{equation}
    3T_\phi n_\phi \braket{C_{h\to \phi S S^* }}_2 \simeq  \frac{\lambda_{\phi s}^2\theta^2}{3072\pi^5}m_h^4 T K_2(m_h/T)\,.
\end{equation}
The result for $S$ is analogous.
\subsection{Higgs annihilation}
\label{sec:Higgs_ann}
The zeroth moment is 
\begin{equation}
     n_\varphi \braket{C_{hh\leftrightarrow \varphi\varphi}} = \frac{1}{2!}\int d\Pi_1 d\Pi_2 d\Pi_3 d\Pi_4 \vert\tilde{\mathcal{M}}_{hh\leftrightarrow\varphi\varphi}\vert^2\left(f_1f_2 -f_3 f_4\right)\,, 
\end{equation}
where we have labeled the momenta as $h_1 h_2\leftrightarrow \varphi_3\varphi_4$. As in the decay case, we will neglect Higgs production. This integral can be cast in terms of the cross section~\cite{Edsjo:1997bg},
\begin{equation}\label{eq:integral_hh_to_phiphi}
      n_\varphi \braket{C_{hh\to \varphi\varphi}}= \frac{T}{32\pi^4}\int_{4m_h^2}^\infty \,ds\,\sqrt{s}(s-4m_h^2)K_1(\sqrt{s}/T)\,\sigma_{hh\to\varphi\varphi}\,.
\end{equation}

On the other hand, the second moment is
\begin{equation}
 3T_\varphi n_\varphi\braket{C_{hh\to\varphi\varphi}}_2 = \frac{1}{2!}\int d\Pi_1\dots d\Pi_4\,\vert\tilde{\mathcal{M}}_{hh\leftrightarrow\varphi\varphi}\vert^2\frac{\vec p_3^{\,2}}{E_3}\,f_1 f_2\,.
\end{equation}
We boost to the CM of mass frame as in eq.~\eqref{eq:ap_CMboost}. In this case $z=E_+/\sqrt{s}$, where $E_+=E_1+E_2$. The two final states integral results in
\begin{equation}
    \begin{split}
        &\int d\Pi_3 d\Pi_4\,\vert\tilde{\mathcal{M}}_{hh\to\varphi\varphi}\vert^2\,\left(E_3 z+\vert \vec p_3\vert\cos\theta_3\sqrt{z^2-1}-\frac{m_\varphi^2}{E_3 z + \vert \vec p_3\vert \cos\theta_3\sqrt{z^2-1}}\right)
        \\&=\frac{\vert\mathcal{M}_{hh\to\varphi\varphi}\vert^2}{8\pi}\frac{\vert\vec p_3\vert}{\sqrt{s}}\left(\sqrt{s}z- \frac{m_\varphi^2}{\vert\vec p_3\vert}\frac{1}{\sqrt{z^2-1}}\log\left(\frac{\sqrt{s} z + \vert\vec p_3\vert\sqrt{z^2-1}}{\sqrt{s} z - \vert\vec p_3\vert\sqrt{z^2-1}}\right)\right)\,.
    \end{split}
\end{equation}
%
\subsection{Production from electroweak states}\label{ap:EW}
The freeze-in contribution involving electroweak states is particularly relevant, as the propagator of gauge bosons remains approximately constant at high energies. We obtain the matrix elements using the \texttt{CalcHEP 3.8.10} package~\cite{Belyaev:2012qa} and employ the high energy limit in the cBE. The matrix elements are quite lengthy, so here we show the cross sections at high $s$ (and $m_\phi \to 0$):
\begin{equation}
    \begin{split}
    \sigma_{hh\to h\phi}&= \frac{\lambda_h^2\theta^2}{16\pi s}\,,
    \\
    \sigma_{Z h\to Z\phi}&= \theta^2 \frac{s-2m_Z^2-6m_h^2+12m_Z^2\log \left( \frac{s}{m_Z^2}\right)}{144\pi v^4}\,,
    \\
    \sigma_{W^{\pm}Z\to W^\pm \phi} &= \theta^2\frac{480m_W^4-72m_W^2m_Z^2 + 24m_Z^4}{864 m_Z^2 \pi v^4}\,,
    \\
    \sigma_{W^{+}W^{-}\to h \phi} &= \theta^2\frac{s-2m_W^2+8m_W^2\log\left(\frac{s}{m_W^2}\right)}{144\pi v^4}\,,
    \\
     \sigma_{W^{+}W^{-}\to Z \phi} &= \theta^2\frac{m_W^2(8m_W^2+m_Z^2)}{18m_Z^2\pi v^4}\,,
     \\
     \sigma_{W^{\pm} h\to W^{\pm} \phi} &=\theta^2\frac{12m_W^2\log \left(\frac{s}{m_W^2}\right)-6m_h^2-2m_W^2+s}{144\pi v^4} \,,
    \\
     \sigma_{ZZ\to h\phi} &=\theta^2\frac{s+3m_h^2 - 2m_Z^2 + 8m_Z^2\log\left(\frac{s}{m_Z^2}\right)}{144\pi v^4}\,.
    \end{split}
\end{equation}

\section{Dark Matter - mediator interactions}\label{ap:C_DM_med}
%
\subsection{Production-annihilation}
The collision operator for the $\phi\phi\leftrightarrow SS^*$ reaction is given by
\begin{equation}
    \begin{split}
        C_{\phi\phi\leftrightarrow SS^*}[S]=\frac{1}{2E_Sg_S}\int\Bigg(&-f_S\vert\tilde{\mathcal{M}}_{\underline S S_2^*\to\phi_3 \phi_4}\vert^2 \,d\Pi_2 f_2\left(\frac{1}{2!}d\Pi_3 d\Pi_4 \right)
        \\&+\vert\tilde{\mathcal{M}}_{\underline S S_2^*\leftarrow\phi_3 \phi_4}\vert^2 d\Pi_2\left(\frac{1}{2!}d\Pi_3d\Pi_4\,f_3f_4\right) \Bigg)\,.
    \end{split}
\end{equation}
Note that we assume that the mediator sector is also diluted. The zeroth moment thermal average reads
\begin{equation}
\begin{split}
   &\braket{C_{\phi\phi\leftrightarrow SS^*}[S]}=\\&-\left(\frac{n_S}{n_S^{\text{eq}}}\right)^2\frac{\lambda_{\phi s}^2}{2!n_S}\frac{T_S}{512\pi^5}\int_{\max(4m_\phi^2,4m_s^2)}^\infty ds\,\frac{\sqrt{s-4m_\phi^2}\sqrt{s-4m_s^2}}{\sqrt{s}}K_1(\sqrt{s}/T_S)
   \\&+\left(\frac{n_\phi}{n_\phi^{\text{eq}}}\right)^2\frac{\lambda_{\phi s}^2}{2!n_S}\frac{T_\phi}{512\pi^5}\int_{\max(4m_\phi^2,4m_s^2)}^\infty ds\,\frac{\sqrt{s-4m_\phi^2}\sqrt{s-4m_s^2}}{\sqrt{s}}K_1(\sqrt{s}/T_\phi)\,.
\end{split}
\end{equation}
In the relativistic regime the lightest particle can be considered essentially massless. In order to obtain the first non-relativistic correction, we expand $\sqrt{s-4m^2}\approx \sqrt{s}-2m^2/\sqrt{s}$, where in this case $m=\min(m_\phi,m_s)$. Hence, we estimate the integral as
\begin{equation}
\begin{split}
     \int_{4M^2}^\infty ds\,\frac{\sqrt{s-4m_\phi^2}\sqrt{s-4m_s^2}}{\sqrt{s}}K_1(\sqrt{s}/T_\text{ds})\approx &4M^2 T_\text{ds} K_1(M/T_\text{ds})^2
     \\&-e^{-2M/T_\text{ds}}\frac{m^2 T_\text{ds}^2}{M}\,,
    \end{split}
\end{equation}
with $M=\max(m_\phi,m_s)$ and $T_\text{ds}$ stands for either $T_\phi$ or $T_S$. Finally, we notice that $\braket{C_{\phi\phi\leftrightarrow SS^*}[\phi]} = -2\frac{n_S}{n_\phi}\braket{C_{\phi\phi\leftrightarrow SS^*}[S]}$. We use this approximation in the cBE, Eq.~\eqref{eq:system_med_DM}.

\subsection{Scattering}
We now compute the collision operator for the scattering between the DM and mediator sectors. We adopt the parametrization of~\cite{Aboubrahim:2023yag},
\begin{equation}
     C_\text{scatter}[S] = \frac{1}{128\pi^3 E_1 g_S\vert\vec p_1\vert}\int_{m_S}^\infty dE_3\int_{\text{max}(m_\phi,E_3-E_1+m_\phi)}^\infty dE_2\,\Pi(E_1,E_2,E_3)\mathcal{P}(f_1,\dots,f_4)\,,
\end{equation}
where we label the momenta as $S_1 \phi_2\leftrightarrow S_3 \phi_4$. The integrand factor $\Pi$ is defined as
\begin{equation}
\Pi(E_1,E_2,E_3) = \lambda_{\phi s}^2\left(k_+ - k_-\right)
\Theta(k_+-k_-)\,,
\end{equation}
with $k_+=\text{min}\left(\vert\vec p_1\vert + \vert\vec p_3\vert, \vert \vec p_2\vert + \vert\vec p_4\vert \right)$, $k_-=\text{max} \left( \left\vert \vert \vec p_1\vert - \vert \vec p_3\vert \right\vert, \left\vert \vert\vec p_2 \vert - \vert \vec p_4\vert \right\vert    \right)$ and $\Theta$ the Heaviside step function. The functional $\mathcal{P}$ incorporates the distribution functions,
\begin{equation}
    \mathcal{P}(f_1,f_2,f_3,f_4)= f_3 f_4 - f_1 f_2\,.
\end{equation}
Since the matrix element is constant, we can estimate the collision operator in the relativistic limit, i.e. we approximate $E_{j}\approx \vert\vec p_{j}\vert$, for $j$ any initial or final state. 
The second moment in the relativistic limit takes the form (setting $g_S=1$)
\begin{equation}
\begin{split}
   \braket{C_\text{scatter}[S]}_2 &\simeq \frac{1}{n_S\,3 T_S}\frac{1}{2\pi^2}\int_0^\infty dE_1\,E_1^2\,C_\text{scatter}
   %
   %
   \\&=\frac{\lambda_{\phi s}^2}{n_S\,3 T_S\,256\pi^5}\int_0^\infty dE_1\,E_1 \int_0^\infty dE_3\int_{\text{max}(0,E_3-E_1)}^\infty dE_2\,\Pi\,\mathcal{P} \,.
\end{split}
\end{equation}
Note that the integrand is
\begin{equation}
    \Pi\,\mathcal{P}=\,(E_1 + E_2 - \vert E_1 - E_3\vert - \vert E_2 - E_3\vert)\Theta(E_1 + E_2 - \vert E_1 - E_3\vert - \vert E_2 - E_3\vert)(f_3 f_4-f_1f_2)\,,
\end{equation}
and
\begin{equation}
    \begin{split}
    f^\text{eq}_3f^\text{eq}_4-f^\text{eq}_1f^\text{eq}_2 = (e^{-E_3/T_S}e^{-(E_1-E_3)/T_\phi}-e^{E_1/T_S})e^{-E_2/T_\phi}\,.
    \end{split}
\end{equation}
We can perform the integral over $E_2$ analytically, for this we split the integral in two cases: $E_2<E_3$ and $E_2>E_3$. The result is:
\begin{equation}
    \begin{split}
        \int_{\text{max}(0,E_3-E_1)}^{E_3} dE_2&\,(E_1 + 2E_2 - E_3 -\vert E_1-E_3\vert  )e^{-E_2/T_\phi} 
        \\&+ \int_{E_3}^\infty dE_2\,(E_1 + E_3 -\vert E_1-E_3\vert)e^{-E_2/T_\phi}
        \\&=2e^{-E_3/T_\phi}T_\phi^2\left(e^{\frac{E_1+E_3-\vert E_1-E_3\vert}{2T_\phi}}-1\right )\,,
    \end{split}
\end{equation}
while the integral over $E_3$ is
\begin{align}
        &2T_\phi^2\int_0^\infty dE_3\,e^{-E_3/T_\phi}\left(e^{\frac{E_1+E_3-\vert E_1-E_3\vert}{2T_\phi}}-1\right )\left(e^{-E_3/T_S}e^{-(E_1-E_3)/T_\phi}-e^{E_1/T_S}\right)\notag\\
        &=2T_\phi^2\frac{e^{-E_1(1/T_\phi+1/T_S)}(e^{E_1/T_S}T_S^2-e^{E_1/T_\phi}\left(E_1(T_\phi-T_S)+T_S^2 \right))}{T_\phi-T_S}\,.
\end{align}
Integrating this result with $\int d E_1\,E_1 $ yields
\begin{equation}\label{eq:S_med_scatter}
    \braket{C_\text{scatter}[S]}_2\simeq \left(\frac{n_S}{n_S^\text{eq}}\right)\left(\frac{n_\phi}{n_\phi^\text{eq}}\right) \frac{\lambda_{\phi s}^2}{n_S\,3T_S\,128\pi^5}T_\phi^2T_S^2(T_\phi-T_S)e^{-m_S/T_S}e^{-m_\phi/T_\phi}\,.
\end{equation}
We introduce the exponential factor to account for non-relativistic effects. To incorporate scattering in $x_\phi'$, we observe that $\braket{C_\text{scatter}[\phi]} = -\frac{3T_S n_S}{3T_\phi n_\phi} \braket{C_\text{scatter}[S]}$.
\section{Vacuum stability}\label{ap:vac_stab}
%
The mediator-Higgs interactions are encoded in the following potential
\begin{equation}\label{V(h,phi)}
V(H,\phi) =  \mu_h^2\,H^\dagger H + \lambda_h(H^\dagger H)^2 +(B_{h\phi}\phi + \lambda_{h\phi}\phi^2)H^\dagger H 
+\lambda_1\phi +\frac{1}{2}\mu_\phi^2\phi^2 + \frac{g_\phi}{3!}\phi^3 + \frac{\lambda_\phi}{4!}\phi^4 \,,
\end{equation}
where $H$ is the $\text{SU}(2)_\text{L}$ Higgs doublet of the SM. Working with the unitarity gauge, $H = \frac{1}{\sqrt{2}}(0,h)^\intercal$; after EWPT, the Higgs boson acquires a VEV, $h\to h+v$, with $v\simeq 246\,\text{GeV}$.  Note that we include the linear term $\lambda_1\phi$. In models with dark scalars, this term is usually neglected by $\lambda_1=0$. The general argument is that we can shift the field $\phi\to\phi+\phi_0$, rearrange terms and demand the resulting factor in the linear term to be zero~\cite{Fradette:2017sdd}.  This is the standard procedure for finding the minima of the potential, as the minimization condition is equivalent to expanding the potential as \(\phi \to \phi + w\) and demanding the linear term to vanish. For now, let us consider \(\lambda_1 \neq 0\). The extrema are given by the solutions of
\begin{equation}\label{extremum_cond}
\begin{split}
\left.\frac{\partial V}{\partial \phi}\right\vert_{h=v,\phi = w} &= \frac{g_\phi}{2}w^2 + \frac{\lambda_\phi}{6}w^3 
+ \mu_\phi^2 w + v^2w\lambda_{h\phi} + \frac{1}{2}v^2 B_{h\phi}+\lambda_1=0\,\qquad\text{and}
\\ \left.\frac{\partial V}{\partial h}\right\vert_{h=v,\phi = w} &= \lambda_hv^3 + v\mu_h^2 + vwB_{h\phi} + vw^2\lambda_{h\phi}=0\,,
\end{split}
\end{equation}
substituting
\begin{equation}\label{lambdas}
\begin{split}
\lambda_h&= -\frac{\mu_h^2-w B_{h\phi}-w^2\lambda_{h\phi}}{v^2}\,\qquad\text{and}
\\ \lambda_1 &= -\left(\frac{g_\phi}{2}w^2 + \frac{\lambda_\phi}{6}w^3 
+ \mu_\phi^2 w + v^2w\lambda_{h\phi} + \frac{1}{2}v^2 B_{h\phi} \right)\,,
\end{split}
\end{equation}
we ensure that the potential at $\braket{\phi}=w$ and $\braket{h} = v \cong 246\,\text{GeV}$ is a critical point. These substitutions also ensure that the linear term (tadpole) vanishes. If $w=0$ is the solution to~\eqref{extremum_cond}, then $v^2 B_{h\phi}/2 + \lambda_1 = 0$. Alternatively, starting with $(v^2B_{h\phi}/2 + \lambda_1)\phi = 0$ implies that one solution for $w$ is $w=0$. For simplicity, we assume that a solution is $w=0$, which implies $v^2 B_{h\phi}/2 + \lambda_1= 0$ (this is the redefinition adopted in the literature~\cite{Fradette:2017sdd}). The first equation in~\eqref{extremum_cond} transforms to
\begin{equation}\label{extremum_cond2}
\frac{g_\phi}{2}w^2 + \frac{\lambda_\phi}{6}w^3 + \mu_\phi^2 w + v^2w\lambda_{h\phi}=0\,.
\end{equation}
The mass matrix is given by the second derivatives of the potential evaluated at $(v,w)$,
\begin{equation}
M^2 = 
\begin{pmatrix}
-2\left(\mu_h^2+w(B_{h\phi}+w\lambda_{h\phi})\right) & v\left(B_{h\phi}+2w\lambda_{h\phi}\right)
\\v\left(B_{h\phi}+2w\lambda_{h\phi}\right) & g_\phi w+\frac{w^2\lambda_\phi}{2}+\mu_\phi^2 + v^2\lambda_{h\phi}
\end{pmatrix}\,,
\end{equation}
%
which can be diagonalized via the rotation matrix
\begin{equation}\label{eq:rot_matrix}
\mathcal{O} = \begin{pmatrix} \cos\theta&\sin\theta\\ - \sin\theta &\cos\theta\end{pmatrix}\,,
\end{equation}
such that $\mathcal{O}^\intercal M^2\mathcal{O} = \text{diag}(m_h^2,m_\phi^2)$. 
The off-diagonal term yields to 
\begin{equation}\label{eq:mixing_angle}
\sin 2\theta  =\frac{2v(B_{h\phi} +2w\lambda_{h\phi})}{m_\phi^2-m_h^2}\,.
\end{equation}

Note that $w \neq 0$ yields the same physics as $w = 0$ due to the singlet nature of $\phi$, as no symmetry is broken when $w \neq 0$. However, it is crucial to consider the additional solutions from~\eqref{extremum_cond2}, which may not necessarily correspond to minima,
\begin{equation}
\begin{split}
w_\pm &= \frac{-3g_\phi\pm \sqrt{3}\sqrt{3g_\phi^2 - 8\lambda_\phi \mu_\phi^2 - 8v^2\lambda_\phi\lambda_{h\phi}}}{2\lambda_\phi}\,.
\end{split}
\end{equation}
If we impose the constraint 
\begin{equation}
0>3g_\phi^2 - 8\lambda_\phi \mu_\phi^2 - 8v^2\lambda_\phi\lambda_{h\phi} \,,
\end{equation}
then the solutions $w_\pm$ do not exist. The last expression is equivalent to
\begin{equation}
3g_\phi^2 - 8\lambda_\phi \mu_\phi^2 - 8v^2\lambda_\phi\lambda_{h\phi}= 3g_\phi^2-4(m_\phi^2+m_h^2)\lambda_\phi + 4(m_h^2-m_\phi^2)\lambda_\phi\cos 2\theta\,.
\end{equation}
Since we are interested in the limit $\theta \ll 1$, the last condition
can be expressed as
\begin{equation}
3g_\phi^2-8m_\phi^2\lambda_\phi<0\implies \frac{3g_\phi^2}{8m_\phi^2}<\lambda_\phi\,.
\end{equation}
We also assume that $\lambda_\phi$ and $g_\phi$ are very small, preventing $\phi$ from chemically thermalizing the mediator sector. Under these assumptions, the minimization of the potential given in Eq.~\eqref{eq:full_potential} yields the solutions $\braket{S} = 0$ (provided $k < \frac{8}{3}$) and $w = 0$.

\clearpage
{\small
\bibliography{biblio.bib}
\bibliographystyle{JHEP}
}





\end{document}